\documentclass[a4paper,11pt]{article} 
\usepackage[margin=0.75in]{geometry} 
\usepackage{amssymb}
\usepackage{amsmath,epsfig, epsf, epic, graphicx} 
\usepackage{fancybox, lscape} 
\usepackage{bm}
\usepackage{natbib}
\usepackage{graphicx}
\DeclareGraphicsExtensions{.eps,.pdf,.jpeg,.png}
\usepackage{caption}
\usepackage[margin=2cm]{caption}
\usepackage{setspace}
\usepackage{cleveref}
\usepackage{xcolor}
\usepackage{array}
\usepackage{wrapfig}
\usepackage{multirow}
\usepackage{url}
\usepackage{tabularx}
\usepackage{csquotes}
\usepackage{amsthm}
\usepackage{amsmath}
\allowdisplaybreaks

\usepackage{floatrow}
\usepackage{subfig}
\newcommand{\indep}{\raisebox{0.05em}{\rotatebox[origin=c]{90}{$\models$}}}

\newtheorem{assumption}{Assumption}
\newtheorem*{assumption*}{Assumption}

\title{Causal inference and racial bias in policing: New estimands and the importance of mobility data}
\author{Zhuochao Huang\footnote{PhD Student, Department of Statistics, University of Florida (Email: zhuochao.huang@ufl.ed)}, Brenden Beck\footnote{Associate Professor, School of Criminal Justice, Rutgers University Newark (Email: brenden.beck@rutgers.edu)} and Joseph Antonelli\footnote{Assistant Professor, Department of Statistics, University of Florida (Email: jantonelli@ufl.edu)}}
\date{}

\begin{document}

\maketitle

\begin{abstract}
Studying racial bias in policing is a critically important problem, but one that comes with a number of inherent difficulties due to the nature of the available data. In this manuscript we tackle multiple key issues in the causal analysis of racial bias in policing. First, we formalize race and place policing, the idea that individuals of one race are policed differently when they are in neighborhoods primarily made up of individuals of other races. We develop an estimand to study this question rigorously, show the assumptions necessary for causal identification, and develop sensitivity analyses to assess robustness to violations of key assumptions. Additionally, we investigate difficulties with existing estimands targeting racial bias in policing. We show for these estimands, and the estimands developed in this manuscript, that estimation can benefit from incorporating mobility data into analyses. We apply these ideas to a study in New York City, where we find a large amount of racial bias, as well as race and place policing, and that these findings are robust to large violations of untestable assumptions. We additionally show that mobility data can make substantial impacts on the resulting estimates, suggesting it should be used whenever possible in subsequent studies.
\end{abstract}

{\it Keywords:}  Causal inference, Mobility data, Racial discrimination, Race and place, Sensitivity analysis.

\section{Introduction}

Understanding the extent to which racial bias can impact policing decisions is a crucially important research question. Due to the relevance of this issue in society today, it has received significant attention with many studies identifying racial bias in a number of distinct criminological areas \citep{antonovics2009new, baumgartner2018suspect, edwards2019risk}. Despite the large body of literature, a coherent conclusion remains elusive as some studies find disparate results \citep{fridell2017explaining}. While there are many potential reasons for these conflicting findings, one such reason is that most researchers have not grappled with the inherent complexities of using administrative data, where the data collected is itself influenced by an individual's race. In our context, issues arise because we only get to observe civilian-police encounters that led to a police stop, but these are not expected to be a random subset of all police encounters, and who is stopped could be a function of racial bias. In this manuscript, we build on this literature by improving on existing estimands and estimation strategies to study racial bias in policing in New York City (NYC) in a way that addresses difficulties with administrative data, and acknowledges uncertainty in whether particular assumptions hold through sensitivity analysis. In addition, we study a secondary question examining the existence of ``race and place" policing, which is the belief that certain racial groups are policed differently based on the racial composition of the neighborhood they are in. Existing evidence suggests racial composition does play a significant role in policing decisions \citep{novak2012racial,meehan2013race,gaston2019enforcing, gaston2019producing, gaston2020minorities, zhang2021racial, chamberlain2022drunk}, however, most of this literature has ignored inherent difficulties with causal inference in this setting caused by selection bias and unmeasured confounding. We develop novel estimands targeting this question, which are robust to certain forms of unmeasured confounding likely to be present in policing studies and address limitations due to the administrative nature of the data. 

Much of the existing empirical work examining questions of racial bias has focused solely on accounting for omitted variable bias, which would occur if a variable associated with both race and the outcome of interest were not included in an analysis. While this issue may be possible in observational studies of racial bias, it is not the only issue to be addressed as there are major issues stemming from selection bias \citep{heckman1979sample} and adjusting for a post-treatment variable \citep{rosenbaum1984consequences, elwert2014endogenous}. Selection bias may occur in our setting if one race is stopped more than another simply due to their race, and not due to other factors or criminologic behavior. One innovative approach to dealing with some of these issues utilizes the so-called ``veil of darkness "\citep{grogger2006testing}, which compares traffic stops before and after dark as officers would not be able to see the race of an individual after dark. Researchers have also utilized outcome tests, which compare success rates of decisions across racial groups as any differences would indicate racial bias \citep{ayres2002outcome, goel2016precinct}. One issue with these tests is that in the absence of discrimination, the detainment rates for minorities and whites might differ if the two groups have different risk distributions of committing crimes \citep{simoiu2017problem, neil2019methodological}. Recent statistical work in \cite{jung2018mitigating} looks at studies of discrimination and highlights how issues can also occur from included variable bias. They develop an approach that addresses this potential source of bias, while also developing a formal sensitivity analysis framework to assess the robustness of their findings to omitted variable biases. It is also worth mentioning that similar statistical issues regarding selection bias and conditioning on post-treatment variables occur in other contexts outside of policing, such as in cluster randomized trials \citep{li2022clarifying, papadogeorgou2023addressing}, studies with truncation by death \citep{ding2011identifiability}, and noncompliance in randomized experiments \citep{angrist1993identification}. Despite this existing literature, policing and studies of racial discrimination come with unique challenges not addressed in these related problems, and therefore we now focus on the recent causal inference literature highlighting issues in this setting in particular. 

Recent work in \cite{gaebler2022causal} formalizes causal inference for studies of discrimination. They focused on the timing of treatment and careful definition of the estimand to allow for estimation of a second stage causal effect, which in our scenario would be the effect of race on police use of force, but only among those encounters that led to a police stop. They utilized sensitivity analysis approaches developed in \cite{cinelli2020making} to assess the robustness of their approach to unmeasured confounding bias. This approach ignores discrimination in the first stage, which is the decision by the police to stop someone in the first place. Given the prior literature suggesting a strong effect of race in this decision \citep{gelman2007analysis}, this is not the estimand of interest in our study as it may underestimate the extent of discrimination, if it exists.  

Another approach to estimating racial disparities can be found in causal decomposition analysis \citep{jackson2018decomposition}, which uses stochastic interventions to assess how much of an observed difference between two groups occurs even when the distribution of an intervention is equalized between the two groups. This framework has been utilized to study the impact of race on medical decisions \citep{ben2024estimating} or the impact of parental support on suicidal ideation \cite{shen2025calibrated}. The latter study additionally developed a framework for sensitivity analysis to unmeasured confounding based on a marginal sensitivity model, which has connections to the sensitivity analyses we propose in subsequent sections. Another work with particular relevance to our study is that of \cite{knox2020administrative}, who discussed many of the difficulties of administrative data in policing research where we only get to observe data among encounters that led to a police stop. They utilized principal stratification \citep{frangakis2002principal} to show that estimands defined over populations in administrative records do not correspond to estimands of interest. They discussed how this can be alleviated with sensitivity analysis along the lines of \cite{manski1990nonparametric} or by incorporating a secondary data source, which can be used to identify crucial associations that are not identifiable in the data set of interest. This work was extended in \cite{zhao2022note} by looking at a ratio-based estimand that relies on weaker assumptions than those in \cite{knox2020administrative}, however this estimand generally involves an unidentified component that requires secondary data to estimate, which comes with inherent difficulties that we discuss in subsequent sections. 

We extend these works in a number of distinct and important directions. First, we utilize anonymized and aggregated cell phone mobility data to improve inference on the causal risk ratio (CRR), which was defined in \cite{zhao2022note}. This estimand requires a secondary data source that effectively identifies the racial composition of the area being policed. The authors in \cite{zhao2022note} use Census data that provides information on racial composition of people living in a police precinct, but this does not reflect the racial composition throughout the day due to travel across precincts. We use anonymized and aggregated cell phone mobility data to obtain an improved estimate of this population and find that it leads to drastically different estimates about the nature of discrimination. We then propose a novel estimand, which targets the question of race and place policing. We show that this estimand is robust to certain types of unmeasured confounding that can affect existing estimands targeting racial discrimination, and we develop a sensitivity analysis procedure assessing the robustness of our estimates to a variety of potential deviations from our core assumptions. We estimate both of these estimands on the data set provided in \cite{fryer2019empirical} to provide a clearer picture of racial discrimination of policing in New York City. Overall, we find that there is significant discrimination in the city, and that Black individuals are policed differently in predominantly white neighborhoods. Importantly, these findings are robust to large violations of identifying assumptions necessary to estimate causal effects in this setting. 

\section{Data structure, notation, and potential outcomes}

Throughout, we assume that our unit of analysis is any encounter between a police officer and a civilian, whether it leads to a stop or not. Our data therefore consists of $(Y_i, M_i, D_i , X_i)$ for $i = 1, \dots, n$. For simplicity, we drop the subscript $i$ when it is not necessary for exposition. We denote $Y_i$ as the observed outcome, which is whether police used force against a civilian in encounter $i$. We let $M_i$ denote whether the encounter led to a police stop. The race of an individual is given by $D_i$ and throughout we let $D_i = 1$ denote Black individuals and $D_i = 0$ denote white individuals. We exclude other races for the purposes of this manuscript, but the same ideas can be applied analogously to compare any two groups. $X_i$ is the set of covariates for encounter $i$ and contains demographic variables of the civilian as well as other features about the encounter such as time of day. It is important to note that our administrative data set consists of all stops in NYC over the years 2003-2013 and therefore we only observe data with $M_i = 1$. We utilize the potential outcome framework \citep{rubin1974estimating} and let $M_i(d)$ represent the potential mediator denoting whether encounter $i$ would have led to a stop if the civilian's racial status were set to $d$. Similarly, $Y_i(d,m)$ represents the potential outcome for the application of force that would have been observed if race was set to $d$ and stop status set to $m$. We also let $Y_i(d) = Y_i(d, M(d))$ be the potential outcome if race was set to $d$. Defining potential outcomes in this manner inherently assumes the Stable Unit Treatment Value Assumption(SUTVA, \cite{rubin1980randomization}), so that $M_i(D_i)= M_i$ and $Y_i(D_i, M_i) = Y_i(D_i) = Y_i$. This assumption links our observed data to the potential outcomes if there are not multiple versions of treatment and there is no interference across units. No interference is not a strong assumption here as it is unlikely that the race of one civilian affects whether another civilian is stopped. Note that we also assume $Y_{i}(d,0) = 0 \text{ for all } i \text{ and for } d \in \{0, 1\}$, which is a very mild assumption as it is reasonable to assume an officer must first detain an individual in order to use force against them. There is debate, however, about defining potential outcomes and causal effects for race or other immutable treatments \citep{pearl2018does}. While some have argued that causal effects of such treatment variables are not well-defined \citep{holland1986statistics}, many others focus on perception of race as a treatment variable \citep{greiner2011causal}. We do not intend to add to this debate, but we focus on defining treatment as the police officer's perception of race in a given encounter, which we believe to be well-defined and lead to coherent causal estimands.

\section{Causal risk ratio and cell phone mobility data}

We first examine the causal risk ratio, which is defined as
$$ \text{CRR}(x) = \frac{E[Y(1) \mid X=x]}{E[Y(0) \mid X=x]}.$$ 
This estimand was introduced in recent work by \cite{zhao2022note}, and it corresponds to the ratio in the probabilities of use of force by the police if a person is Black compared with being white. It captures discrimination both in the decision of the police to stop an individual, as well as whether to use force against an individual given that they have been stopped. For more information about this estimand, along with its identifying assumptions, we point readers to \cite{zhao2022note}, though a brief review can also be found in Appendix D. We focus here on how anonymized and aggregated cell phone mobility data can help estimate this quantity. Note that in Section \ref{sec:NYPDanalysis} we estimate this quantity separately by precinct, so effectively we have $\text{CRR}_j(x)$ for all precincts $j$, though we suppress dependence on $j$ in what follows for simplicity, but all terms are estimated at the precinct level. Under standard assumptions described in Appendix D, it was shown that this estimand is identified as a function of observed data under the following formulation:
\begin{align*}
    \frac{E[Y(1) \mid X=x]}{E[Y(0) \mid X=x]} &= 
    \frac{E(Y \mid M=1, D=1, X=x)}{E(Y \mid M=1, D=0, X=x)} \\
    &\times \frac{P(D=1 \mid M=1, X=x)}{P(D=0 \mid M=1, X=x)} \\
    &\times \frac{P(D=0 \mid X=x)}{P(D=1 \mid X=x)}
\end{align*}

It is clear from this expression that estimation of CRR$(x)$ relies on estimating three distinct models: 1) $E(Y \mid M=1, D=d, X=x)$, 2) $P(D=d \mid M=1, X=x)$, and 3) $P(D=d \mid X=x)$. The first two of these models are straightforward and can be estimated using generalized linear models or more complex machine learning methods. Both of these models condition on $M=1$ and therefore we can use our observed sample to estimate these quantities. The third quantity does not condition on $M=1$ and therefore can not be estimated using our data without overly strong assumptions. 

Estimation of $P(D=d \mid X=x)$ is problematic because it is a probability in the population of all police-civilian encounters. Not only do we not have data from this population in our setting, but it is inherently a difficult population to ever observe data from directly. To deal with this, one can make approximations or assumptions in conjunction with a secondary data source for estimation. As an example, in prior work \citep{zhao2022note}, the United States Census data was used to estimate this ratio as this data source provides racial demographics across NYC. Using this data source comes with two distinct, and potentially large problems. For one, the Census data provides marginal racial demographics across the city, but does not provide these racial proportions conditional on $X$, which is required for estimating the CRR. They implicitly assumed that $P(D=d \mid X=x) = P(D=d)$, which is unlikely to hold as we can't reasonably assume that race is independent of each observed covariate. The second issue is that Census data estimates marginal demographics of residents who live in each geographic area but does not contain information about the actual population present throughout the day who can interact with police. The racial composition of an area during the day could be significantly different than of the population who live there \citep{athey2021estimating, sampson2023comparative}, which would bias our estimates of this probability and of the CRR estimand. 

\subsection{Improved estimation of $P(D=d \mid X=x)$ using cell phone mobility data}

Addressing the first of these two issues would require a secondary data source from the population of all police-civilian encounters containing information on both $D$ and $X$. Since such a data source is not available (nor likely ever will be available), we instead use marginal estimates of the racial composition of the city, coupled with an assumption that we can perform sensitivity analysis on. Instead of assuming $P(D=d \mid X=x) = P(D=d)$, we make the following assumption:
\begin{assumption}[CRR race ratio assumption]
    \begin{align*}
        \frac{P(D=d \mid X=x)}{P(D=d \mid X=x, M=1)} = \frac{P(D=d)}{P(D=d \mid M=1)}.
    \end{align*}
\end{assumption}
This states that while the ratio of individuals of race $d$ might differ in the full population compared
with the $M=1$ population, this ratio should be the same whether we condition on $X=x$ or not. Re-arranging the terms in this assumption, we see that we can identify $P(D=d \mid X=x)$ as
$$P(D=d \mid X=x) = \frac{P(D=d \mid X=x, M=1) \ P(D=d)}{P(D=d \mid M=1)},$$
all of which are identifiable from either our main administrative data set, which conditions on $M=1$, or the secondary data that provides information on the marginal racial composition of the city. Now that we have the probabilities of interest in terms of quantities we can estimate, we can discuss how to provide better estimates of $P(D=d)$. 

While the Census data provides estimates of $P(D=d)$ for each precinct, they do not accurately reflect the proportion of Black individuals who could potentially interact with police in that precinct due to movement of individuals across precincts. Let $\pi_j$ be the true value of $P(D=d)$ for precinct $j$. Further, let $p_j$ be the residential proportion of individuals who live in police precinct $j$ with $D=d$. In other words, $p_j$ is the estimate of $P(D=d)$ we would use for the CRR if we were only using the Census data. If we let $\boldsymbol{\pi}$ and $\boldsymbol{p}$ represent vectors of these quantities for all precincts, then we can define $$\boldsymbol{\pi} = \boldsymbol{T} \boldsymbol{p}$$
where $\boldsymbol{T}$ is a matrix describing the movement of individuals across the city. Letting $N_{ij}$ be the amount of time that anonymous cell phone users who live in precinct $i$ spend in precinct $j$, then we can estimate
$$\widehat{T}_{ij} = \frac{N_{ij}}{\sum_{k} N_{ik}}.$$
This construction leads to estimates of $\pi_j$ that are weighted averages of the proportion of people with $D=d$, weighted by how much time they spend in precinct $j$. Prior approaches, which utilized Census data only, effectively used $\boldsymbol{T} = \boldsymbol{I}_q$, where $q$ is the number of precincts. As a sensitivity analysis they explored assuming that the police precincts were made up 90\% of people from the home precinct and 10\% from the city-wide average. This amounts to assuming that $\boldsymbol{T} = 0.9 \boldsymbol{I}_q + 0.1 (1/q) \boldsymbol{1}_{q \times q}$. Neither of these is necessarily reflective of the true population in each precinct and can therefore lead to misleading estimates of the CRR, but our estimates based on mobility target the population actually exposed to encounters with the police in a particular precinct. 

The aggregated mobility data used for this study consists of a sample derived from anonymous cell phone users who consented to data collection for research purposes. At no point did researchers attempt to disaggregate data or link mobility records to individuals or individual police stops.  

\section{Race and place estimands}

We now turn attention to a related, but distinct, issue that targets a very specific form of discrimination in police behavior that we refer to as race and place policing. This is the idea that a Black individual is policed differently if they are in a predominantly white neighborhood.

\subsection{Assumptions and definition of estimand}

Before discussing estimands and assumptions, we must define an additional variable, which we denote by $R_i$. We define $R_i$ to be a variable that denotes the proportion of Black individuals who live in the precinct where encounter $i$ took place. Race and place policing implies that whether a Black individual is in a precinct with more or less white individuals can change their probability of being detained and ultimately arrested. Therefore an estimand that targets race and place policing can be defined as
$$\Psi(r, x) = \frac{E[Y(1) \mid R=r, X=x]}{E[Y(1) \mid R=r_0, X=x]},$$
where $r_{0}$ is a baseline value that can be any value in the range of $R$. Note that this choice should not affect the overall findings, however, we express our estimand in this ratio scale relative to $r_{0}$ as it allows for the cancellation of certain probabilities that are difficult to identify with administrative data, leading to less restrictive identification assumptions. Note that unlike for the CRR, we target a city-wide $\Psi(r, x) $, rather than a precinct-specific estimand, as $R$ does not vary among encounters within a single precinct. 

In order to identify this estimand, we first make an ignorability assumption that assumes away certain types of unmeasured confounding.

\begin{assumption}[Race and place ignorability] \label{ass:RacePlaceIgnore}
    \begin{align*}
        &(i)\frac{E[Y(1,1) \mid R=r, X=x, M(1) = 1]}{E[Y \mid R=r, X=x, M = 1, D=1]} = c_1(x) \\
        &(ii)\frac{P(M(1)=1 \mid R=r, X=x)}{P(M=1 \mid R=r, X=x, D=1)} = c_2(x).
    \end{align*}
\end{assumption}

\noindent Here $c_1(x)$ and $c_2(x)$ are unknown functions that do not depend on $r$. Note that this is generally a weaker version of standard ignorability assumptions. Typically, we would have to assume there are no unmeasured confounders of either the $M-D$ or $D-Y$ relationships, such as in Assumption 2, and that $c_1(x) = c_2(x) = 1$. Here we allow for certain types of unmeasured confounding. Part (i) of the assumption could hold in the presence of unmeasured confounding if certain races exhibit more aggression towards police when stopped, resulting in increased forces being used. As long as this increase remains constant irrespective of the value of $r$, the estimand is still identified. Similarly for part (ii) of the assumption, it could be that individuals of one race commit crimes at different rates than those of other races and consequently get stopped at different rates, but this is independent of $r$, as they commit crimes at similar rates across the entire city. Despite the fact that these assumptions are weaker, they are still potentially strong in their own right, and therefore we develop sensitivity analyses for them in Section \ref{ssec:Sensitivity}. 

Under Assumption \ref{ass:RacePlaceIgnore}, we show in Appendix A that this estimand can be written as the following:
\begin{align}
    \Psi(r,x) &= \frac{E(Y \mid R=r, M=1, D=1, X=x)}{E(Y \mid R=r_0, M=1, D=1, X=x)} \nonumber \\
    &\times \frac{P(D=1 \mid R=r, M=1, X=x)}{P(D=1 \mid R=r_0, M=1, X=x)} \nonumber \\
    &\times \frac{f(r \mid M=1, X=x)}{f(r_0 \mid M=1, X=x)} \nonumber \\
    &\times \frac{f(r_0 \mid D=1, X=x)}{f(r \mid D=1, X=x)}, \label{eq:estimate}
\end{align}

\noindent where we adopt the convention that $f(r \mid X=x)$ is the conditional density of $R$ given $X$ evaluated at $r$. We see that this quantity relies on four distinct conditional distributions. Three of these are conditional on $M=1$ and can therefore be estimated using generalized linear models or machine learning techniques on the observed administrative data. As with the CRR, we have a term, which in this case is $f(r \mid D=1, X=x)$, that is not identifiable from the main data set and requires a secondary data source to estimate.

\subsection{Estimation of $f(r \mid D=1, X=x)$ with cell phone mobility data}

As before with the causal risk ratio, we have an unidentified term that does not condition on $M=1$, but our secondary data sources used to estimate it do not provide information on $X$. Specifically, we use a combination of Census data and anonymous cell phone mobility data to estimate $f(r \mid D=1, X=x)$, but given that these do not contain covariate information, we make the following assumption.
\begin{assumption}[Race and place ratio assumption] \label{ass:RacePlaceRatio}
$$\frac{f(r \mid D=1)}{f(r \mid D=1, X=x)} = \frac{f(r \mid D=1, M=1)}{f(r \mid D=1, X=x, M=1)}$$
\end{assumption}
This assumption assumes that while conditioning on covariates $X$ may change the density of $R$ given $D=1$, this ratio should remain constant whether in the stopped population ($M=1$) or in the overall population of interest. For example, Black individuals with covariate value $x$ may be more likely to travel into predominantly white precincts, but this is true both in all encounters with the police, as well as those that led to stops. Re-arranging terms shows that the quantity of interest can be written as
\begin{align*}
    &f(r \mid D=1, X=x) \\
    &= \frac{f(r \mid D=1) \ f(r \mid D=1, X=x, M=1)}{f(r \mid D=1, M=1)}
\end{align*}

The two terms that condition on $M=1$ can be estimated using our main data source in the stopped population, so all that is left is to estimate $f(r \mid D=1)$. 

As before, we utilize anonymized and aggregated cell phone mobility data to better understand the racial composition of individuals within each precinct in order to estimate this probability. We first decompose the remaining term using Bayes rule as

$$f(r \mid D=1) = \frac{P(D=1 \mid R=r) f(r)}{P(D=1)}.$$

First, let's see how to estimate $P(D=1 \mid R=r)$. As discussed above, we estimate $P(D=1)$ for each precinct using the anonymized and aggregated cell phone mobility data. Suppose for precinct $j$, we have an estimate $\widehat{\pi}_j$ of this probability, as well as the proportion of Black individuals in that precinct, which we denote by $r_j$. We use a kernel smoothing approach to estimate $P(D=1 \mid R=r)$ as

\begin{align*}
    \widehat{P}(D=1 \mid R=r) = \frac{\sum_{j} K(r-r_j) \widehat{\pi}_j}{\sum_{j} K(r-r_j)},
\end{align*}
where $K(\cdot)$ is an appropriately chosen kernel that gives larger weights to small values of $r-r_j$. We adopt a similar strategy for $f(r)$. Note that this quantity represents the density of $R$ in the population of all encounters between police and civilians. Because of this, we must incorporate mobility data to account for the fact that people spend differing amounts of time in each precinct. Let $C_j$ be the amount of time spent in precinct $j$ among all anonymous cell phone users. To estimate $f(r)$ we utilize a weighted kernel density estimator defined by
$$\widehat{f}(r) = \frac{\sum_j K(r - r_j) C_j}{\sum_j C_j},$$
where again $K(\cdot)$ is an appropriately chosen kernel that gives larger weights to small values of $r-r_j$. This estimator stems from the fact that every encounter in precinct $j$ has value $R=r_j$, and the estimated density for $R$ upweights precincts that are expected to have more encounters, which we measure through mobility data. 

\section{Challenges with marginalization}
\label{sec:Marginalization}
All of the estimands so far have been conditional estimands, which are functions of a particular value of the covariates $X$. In many settings, it is desirable to produce a single estimand corresponding to the degree of discrimination or race and place policing that is occurring. This requires marginalizing estimands that are functions of $X$ over the distribution of $X$, though there are multiple options for doing this. One natural choice is to proceed with $E_x[\Psi(r, x)]$, which averages the estimand over the distribution of the covariates. One issue with this estimand, which is an issue for all marginal estimands in this setting, is that the marginal distribution of $X$ in the target population of all police-civilian encounters is not known. We only get to observe the distribution of $X$ for encounters that end in a stop, i.e. $M=1$. For this reason, we focus on 
$$\overline{\Psi}(r) = E_{x | M=1}[\Psi(r, X)] \approx \frac{1}{n} \sum_{i=1}^n \Psi(r, X_i).$$
It is important to emphasize that this is still a meaningful estimand that targets causal effects in the full population, not just the $M=1$ population. Local causal effects, as described in \cite{zhao2022note} and \cite{knox2020administrative}, condition on $M=1$ in the definition of the estimand. As such, they necessarily target racial bias—or race and place-based policing—\textit{given that an individual is stopped}, and thus ignore potential racial bias in the initial decision to stop someone. Our marginal estimand here still targets both potential sources of race and place policing, but assigns more weight to covariate values that are more likely to be observed in the $M=1$ population than the full population. This is done to provide a single and interpretable measure of race and place policing, rather than a different estimand for each possible $X$ value, which is difficult to summarize. Additionally, this facilitates sensitivity analyses developed in subsequent sections that focus on marginal estimands. 

Ideally, however, we would have access to the covariate distribution in the full population, and we would implement marginalization using $E_x[\Psi(r, x)]$. This is not identifiable, but we can perform sensitivity analyses to assess the extent to which marginalizing over the wrong covariate distribution could alter our conclusions. If we let $f(x)$ and $f(x \mid M=1)$ be the covariate distributions in the overall and $M=1$ populations, respectively, then we can assume that
$$
\frac{1}{\gamma} \leq \frac{f(x)}{f(x \mid M=1)} \leq \gamma \quad \text{for all } x.
$$
For any value of $\gamma > 1$, we can derive bounds for $E_x[\Psi(r, x)]$, the causal effect that marginalizes over the correct covariate distribution. We provide technical details of this sensitivity analysis, along with its application to the NYPD stop data in Appendix F. In summary, we find that our results are not largely impacted by the marginalization weights, and that the overall conclusions remain unchanged under large values of $\gamma$. 

\section{Sensitivity analyses and bounds for $\Psi(r,x)$}

When studying the race and place estimand, our inferences rely on both an ignorability assumption (Assumption \ref{ass:RacePlaceIgnore}) as well as an assumption on the secondary data source needed to estimate unidentified terms (Assumption \ref{ass:RacePlaceRatio}). In this section we detail approaches to evaluating robustness of our estimates to violations of these assumptions. Additionally, while emphasis in this section is on $\Psi(r,x)$, similar ideas hold for the CRR as well. We focus on $\Psi(r,x)$ as sensitivity analysis and bounds for this estimand are more complicated than for the CRR due to the functional nature of the estimand and the nonstandard ignorability assumption.

\subsection{Sensitivity analysis to violations of assumptions}
\label{ssec:Sensitivity}

In this section we develop a novel sensitivity analysis framework that applies to violations of Assumptions \ref{ass:RacePlaceIgnore}(i), \ref{ass:RacePlaceIgnore}(ii), and \ref{ass:RacePlaceRatio}. While the framework we present is applicable to any of these assumptions, for exposition and building intuition for our sensitivity analysis, we focus specifically on Assumption \ref{ass:RacePlaceIgnore}(i) in what follows. Assessing sensitivity to the other assumptions follows analogously, and we present specific mathematical details for those other assumptions in Appendix E. 

Throughout, we are assuming that this assumption is violated due to the presence of an unmeasured confounder, which we refer to as $U$. We let $\Psi(r, x)$ represent the value we obtain by applying our identification formula in \eqref{eq:estimate} only adjusting for $X$ and ignoring $U$, which does not necessarily equal the true causal effect due to the presence of $U$. We denote the true causal effect by $\Psi^*(r, x)$, which correctly adjusts for both $X$ and $U$. Our sensitivity analysis relies on a key result regarding the differences between these two quantities. First, we can write
    \begin{align*}
        \Psi^*(r, x) &= \frac{E(Y(1) \mid R=r, X=x)}{E(Y(1) \mid R=r_0, X=x)} \\
        &=  \Psi(r, x) \ \xi(r, x),
    \end{align*}
where the form of $\xi(r, x)$ depends on which assumptions are violated due to the presence of $U$, though again we focus on Assumption \ref{ass:RacePlaceIgnore}(i) here. We show in Appendix E that $\xi(r, x)$ takes the form
{\small
\begin{align*}
\xi(r, x) = \left\{ \frac{E_{U \mid R=r, X=x, M(1)=1} \bigg[ \frac{f(U \mid R=r, M=1, D=1, X=x, Y=1)}{f(U \mid R=r, M=1, D=1, X=x)} \bigg]}{E_{U \mid R=r_0, X=x, M(1)=1} \bigg[ \frac{f(U \mid R=r_0, M=1, D=1, X=x, Y=1)}{f(U \mid R=r_0, M=1, D=1, X=x)} \bigg]} \right\}.
\end{align*}
}
This helps build intuition for when the confounding bias from $U$ is expected to be small or not. One can show that $\xi(r, x) = 1$ for all $x$, and therefore there is no bias due to $U$, if any one of three conditions hold:
\begin{enumerate}
    \item $U \indep  Y \mid X, R, M=1, D=1$
    \item (a) $U \indep  R \mid X, M=1, D=1, Y=1$, (b) $U \indep  R \mid X, M=1, D=1$, and (c) $U \indep  R \mid X, M(1) = 1$
    \item $f(U \mid R, X, M(1) = 1) = f(U \mid R, X, M=1, D=1)$
\end{enumerate}
The first two of these conditions show that $U$ must be conditionally associated with both $R$ and $Y$ in order to bias the effect of interest. While less obvious, the third condition effectively shows that $U$ must also be conditionally associated with $D$ given ($R, X, M(1) = 1$) as we detail in Appendix E. This shows that $U$ must affect three different variables in order to obtain bias, unlike standard estimands in which confounders need only be common causes of two variables to lead to bias. Further, the ability to write the true estimand in this form allows us to derive an analytic form for the bias of any marginal estimands which average over the distribution of $X$, such as those described in Section \ref{sec:Marginalization}. We show in Appendix E that the bias of marginal estimands can be written as
\begin{align*}
    & E(\Psi(r, x) - \Psi^*(r, x)) \\
    =& E(\Psi(r, x)(1 - \xi(r, x))) \\
    =& -\text{Corr}(\Psi(r, x), \xi(r, x)) \sqrt{\text{Var}(\Psi(r, x)) \text{Var}(\xi(r, x))} \\
    +& E(\Psi(r, x)) (1 - E(\xi(r, x))),
\end{align*}
where all moments are with respect to the distribution of $X$ that is used during marginalization. One can see that our bias depends on three unidentified parameters: 1) $\text{Corr}(\Psi(r, x), \xi(r, x))$, 2) $\text{Var}(\xi(r, x))$, and 3) $E(\xi(r, x))$. We can use these three as sensitivity parameters to see how much our estimated effects change as a function of these sensitivity parameters. Additionally, because correlations are bounded between -1 and 1, we can provide conservative bounds on the bias that depend only on two sensitivity parameters: the mean and variance of $\xi(r, x)$.  This is the strategy that we pursue through most of the manuscript as it is  easier to reason about two sensitivity parameters than three, and our benchmarking procedure will focus on these two as well, though in principle could be extended to include benchmarking of the correlation parameter.

Note that our sensitivity analysis framework is related to, but distinct, from a recent series of works on sensitivity analysis for average treatment effects for inverse probability weighted estimators \citep{hong2021did,huang2024sensitivity, huang2025variance}. These papers examine the impact of unmeasured confounding on inverse probability weights and develop a sensitivity analysis framework that depends on the variance of the differences between the true and estimated weights, as well as the correlation between these differences and the outcome. Unlike our approach, they do not have a third sensitivity parameter that depends on the means of these differences, because IPW weights have a nice property that guarantees these differences to be zero on average. An additional complexity that we have to grapple with in our sensitivity analysis framework is that we have different sensitivity parameters at each value of $r$ along our functional estimand. We know that $\xi(r_0, x) = 1$ for all $x$, and that $\xi(r, x)$ is generally closer to 1 when $|r - r_0|$ is small, but it is not clear how to specify the sensitivity parameters as an increasing function of $|r - r_0|$. This is one area where benchmarking with observed covariates can be particularly helpful, though we show in the following section that benchmarking with observed covariates comes with unique difficulties for our estimand. Lastly, the sensitivity parameters are arguably less interpretable than others used in the causal inference literature, such as those based on partial $R^2$ values that are easy for researchers to reason about \citep{imbens2003sensitivity, small2007sensitivity, cinelli2020making, chernozhukov2022long}. Having a closed-form expression for $\xi(r, x)$ is somewhat helpful in this regard as we can see the bias depends on imbalance in the unmeasured confounder between data with $Y=1$ and $Y=0$, and how this imbalance differs between $R=r$ and $R=r_0$. Nonetheless, reasoning about the mean and variance of this quantity over the distribution of $X$ can be difficult, which again is a major motivating factor for developing the benchmarking procedure in the following section. Additionally, in Section \ref{ssec:Interpretation} we utilize a simple simulation study to show how these sensitivity parameters vary with parameters that researchers are more accustomed to reasoning about, such as logistic regression parameters. 

\subsection{Benchmarking sensitivity parameters with observed covariates}
\label{ssec:Benchmark}

In this section we describe how to use observed covariates to reason about plausible values of the sensitivity parameters. The core idea behind benchmarking is that one can leave out an observed covariate and calculate the corresponding sensitivity parameters as if the observed covariate were an unmeasured confounder. This can provide us with plausible values of the sensitivity parameters if we are willing to assume that the potential unmeasured confounder's impact is similar to (or some multiple of) the impact of the observed covariate that is left out. While a commonly used strategy in sensitivity analysis, this is particularly challenging for our estimand. The core problem can be seen in the definition of $\xi(r, x)$, where one must calculate an expectation over $U \mid R=r, X=x, M(1) = 1$. If we were to try and benchmark this quantity, it would require taking the expectation over $X_j \mid R=r, X_{-j} = x_{-j}, M(1) = 1$, where we let $X_{-j}$ be the vector of covariates excluding covariate $j$. We do not get to observe $M(1)$ for all observations, and therefore we cannot calculate this expectation. 

As an alternative and feasible path forward, we can first define a different true value of the estimand. If we let $Z = [X, U]$, then define
\begin{align*}
        \widetilde{\Psi}(r, z) &= \frac{E(Y(1) \mid R=r, X=x, U=u)}{E(Y(1) \mid R=r_0, X=x, U=u)} \\
        &=  \Psi(r, x) \ \widetilde{\xi}(r, z),
\end{align*}
The true estimand $\widetilde{\Psi}(r, z)$ can differ from $\Psi(r, x)$ for two main reasons: 1) unmeasured confounding by $U$, and 2) heterogeneity in the effect of $R$ on $Y(1)$ that is driven by $U$. This shows that even if there is no unmeasured confounding by $U$, the value of $\widetilde{\xi}(r, z)$ may not be 1. In Appendix E, we show that this quantity takes the form
\begin{align*}
    \widetilde{\xi}(r, z) &= \frac{f(u \mid R=r, M=1, D=1, X=x, Y=1)}{ f(u \mid R=r, M=1, D=1, X=x)} \\
    & \times \frac{f(u \mid R=r_0, M=1, D=1, X=x)}{f(u \mid R=r_0, M=1, D=1, X=x, Y=1)}
\end{align*}
We can see that this is the same as $\xi(r, x)$ defined above, except it does not include the outer expectations which average over the distribution of $U$. Generally speaking this averaging step will reduce the mean and variance of $\xi(r, x)$ relative to $\widetilde{\xi}(r, z)$. Further, three conditional dependencies must hold for $\xi(r, x)$ to be different from 1, but only the first two of those dependencies need to hold for $\widetilde{\xi}(r, z)$ to deviate from 1. Because of this, we typically see that the sensitivity parameters found when examining $\widetilde{\xi}(r, z)$ are substantially larger than those for $\xi(r, x)$. So while we are not able to benchmark the $\xi(r, x)$ sensitivity parameters, if we can benchmark based on $\widetilde{\xi}(r, z)$, then it is expected we will get conservative results in our sensitivity analysis, further supporting our conclusions if we find that our conclusions are robust to moderate amounts of unmeasured confounding. For this reason, we now proceed to work towards benchmarking $\widetilde{\xi}(r, z)$, which we will use to guide our sensitivity parameters of interest, under the assumption that they will generally lead to more conservative bounds and larger sensitivity parameters. 

A natural first step to benchmarking with this new quantity is to calculate the following:
\begin{align*}
    \widetilde{\xi}_j(r, x) &=  \frac{f(x_j \mid R=r, M=1, D=1, X_{-j}=x_{-j}, Y=1)}{f(x_j \mid R=r, M=1, D=1, X_{-j}=x_{-j})} \\
    &\times \frac{f(x_j \mid R=r_0, M=1, D=1, X_{-j}=x_{-j})}{f(x_j \mid R=r_0, M=1, D=1, X_{-j}=x_{-j}, Y=1)}
\end{align*}
Benchmarking could then proceed by assuming that 
\begin{align*}
    k_m = \frac{E(\widetilde{\xi}(r, z))}{E(\widetilde{\xi}_j(r, x))} \quad \quad k_v = \frac{Var(\widetilde{\xi}(r, z))}{Var(\widetilde{\xi}_j(r, x))},
\end{align*}
and we could posit values of $k_m$ and $k_v$ representing how many times worse the unmeasured confounder is compared with observed covariate $j$ in order to do our sensitivity analysis. Recent work in \cite{cinelli2020making} referred to this type of approach as informal benchmarking, and they showed potential pitfalls of benchmarking in this manner. We show in a simple example in Appendix H that our sensitivity parameters suffer similar issues when doing informal benchmarking. Specifically, the researcher can have correct prior beliefs about the degree of confounding stemming from $U$ compared with $X_j$, yet informal benchmarking underestimates the bias stemming from omitting $U$.

The problem with informal benchmarking is that different sets of variables are conditioned on when calculating $\widetilde{\xi}(r, z)$ and $\widetilde{\xi}_j(r, x)$. Formal benchmarking would require us to posit values of $k_m$ and $k_v$ as above, but where the quantities in both the numerator and denominator condition on the same set of variables, which in this case would be $X_{-j}$. Then we would have to derive how our sensitivity parameters of interest, which condition on the full vector $X$, can be written in terms of $k_m$, $k_v$ and potentially other observable quantities. Unfortunately, no such closed-form expression exists, and therefore we propose a two-step approach to perform an approximate version of formal benchmarking. While our sensitivity parameters $E(\widetilde{\xi}(r,z))$ and $Var(\widetilde{\xi}(r,z))$ are not themselves partial $R^2$ values, they do depend on certain partial $R^2$ values. As the partial $R^2$ between the unmeasured confounder and all of $(D, R, Y)$ increase, our sensitivity parameters generally increase as well. Therefore, as a first step, we utilize the formal bounds derived in \cite{cinelli2020making} to formally benchmark all partial $R^2$ values of interest. As a second step, we take the partial $R^2$ values obtained from formal benchmarking, and we create a $U$ variable that is a function of $(D, R, Y)$ in a way that it satisfies the aforementioned partial $R^2$ values. We can then use this value of $U$ to calculate $\widetilde{\xi}(r, z)$ and obtain $E(\widetilde{\xi}(r,z))$ and $Var(\widetilde{\xi}(r,z))$. Simply knowing the partial $R^2$ values is not sufficient, as the direction of associations also plays a role, so we explore all possible directions of associations between $U$ and the observed variables, and we keep the sensitivity parameters that lead to the largest amount of bias. Full technical details of how to implement formal benchmarking can be found in Appendix G. Overall, we find this procedure works well with respect to benchmarking and show in Appendix H that it addresses the issues caused by informal benchmarking.

\subsection{Interpretation of sensitivity parameters}
\label{ssec:Interpretation}

In this section, we aim to provide intuition about plausible magnitudes of our two sensitivity parameters by relating them to standard regression coefficients when Assumptions \ref{ass:RacePlaceIgnore} or \ref{ass:RacePlaceRatio} fail. Given that most researchers are comfortable reasoning about magnitudes of logistic regression coefficients and odds ratios, this provides an improved understanding of the strength of confounding required to make our sensitivity parameters large. We first focus on violations of Assumption \ref{ass:RacePlaceIgnore} where an unmeasured confounder biases our effects, and then turn to violations of Assumption \ref{ass:RacePlaceRatio} where there are fundamental differences between the $M=1$ and full population. Also, throughout we consider $r_0$ and $r$ to be the first and third quartiles of $R$, respectively, in order to ensure we're evaluating an estimand corresponding to a large shift in $R$. If the difference in $|r - r_0|$ were small, then the sensitivity parameters would be small regardless of the degree of violations of our assumptions.

\subsubsection{Violations of Assumption \ref{ass:RacePlaceIgnore}}

In this section, we present a simple simulation study where an unmeasured confounder affects all variables in $(R, D, M, Y)$ and therefore violates Assumptions \ref{ass:RacePlaceIgnore}(i) and \ref{ass:RacePlaceIgnore}(ii). We leave full details of the data generating processes to Appendix K, though we review the important details here. All binary variables are generated from logistic regression models that are linear in the unmeasured confounder, and $R$ is generated from a linear regression model that is linear in the unmeasured confounder, though we scale $R$ to be between 0 and 1. We assume the same coefficient for the unmeasured confounder in the models for all four of the other variables, and then we increase the value of this coefficient on a grid from 0 to 1. The sign of the coefficient within each model also affects the sensitivity parameters, so we explore all possible signs and keep the largest values of the sensitivity parameters for each size of the coefficient to be varied. Our goal is to see how $E(\xi(r, x))$ and $Var(\xi(r, x))$ change as we increase the regression parameters from no confounding to strong confounding. We also examine $E(\widetilde{\xi}(r, z))$ and $Var(\widetilde{\xi}(r, z))$, which is the estimand that we are able to benchmark as described in the previous section. For our benchmarking procedure to be valid, we need for the sensitivity parameters from $\widetilde{\xi}(r, z)$ to be as large, or larger, than those from $\xi(r, x)$. 

\begin{figure*}[h]
    \centering
    \includegraphics[width=0.9\linewidth]{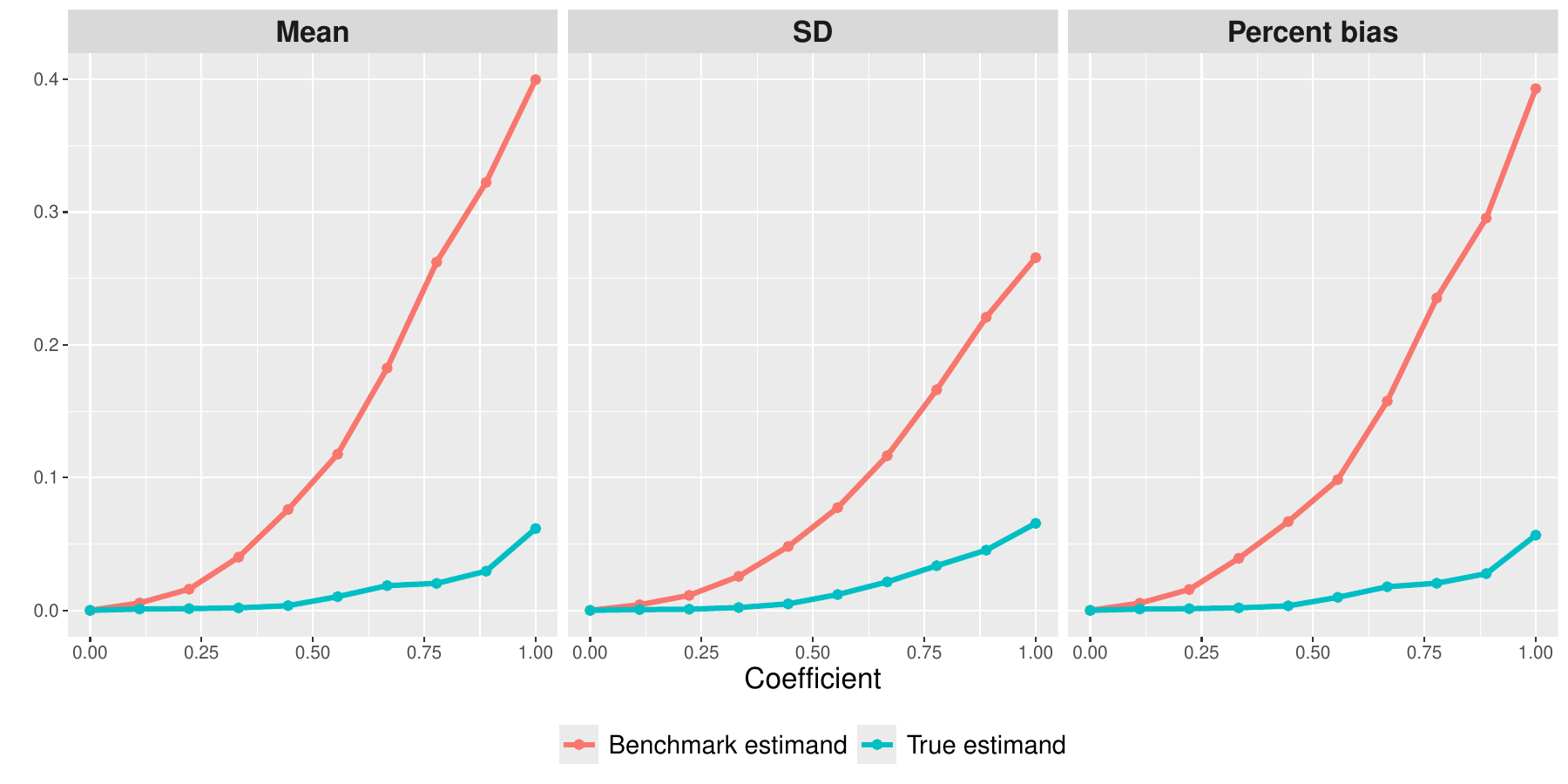} 
    \includegraphics[width=0.9\linewidth]{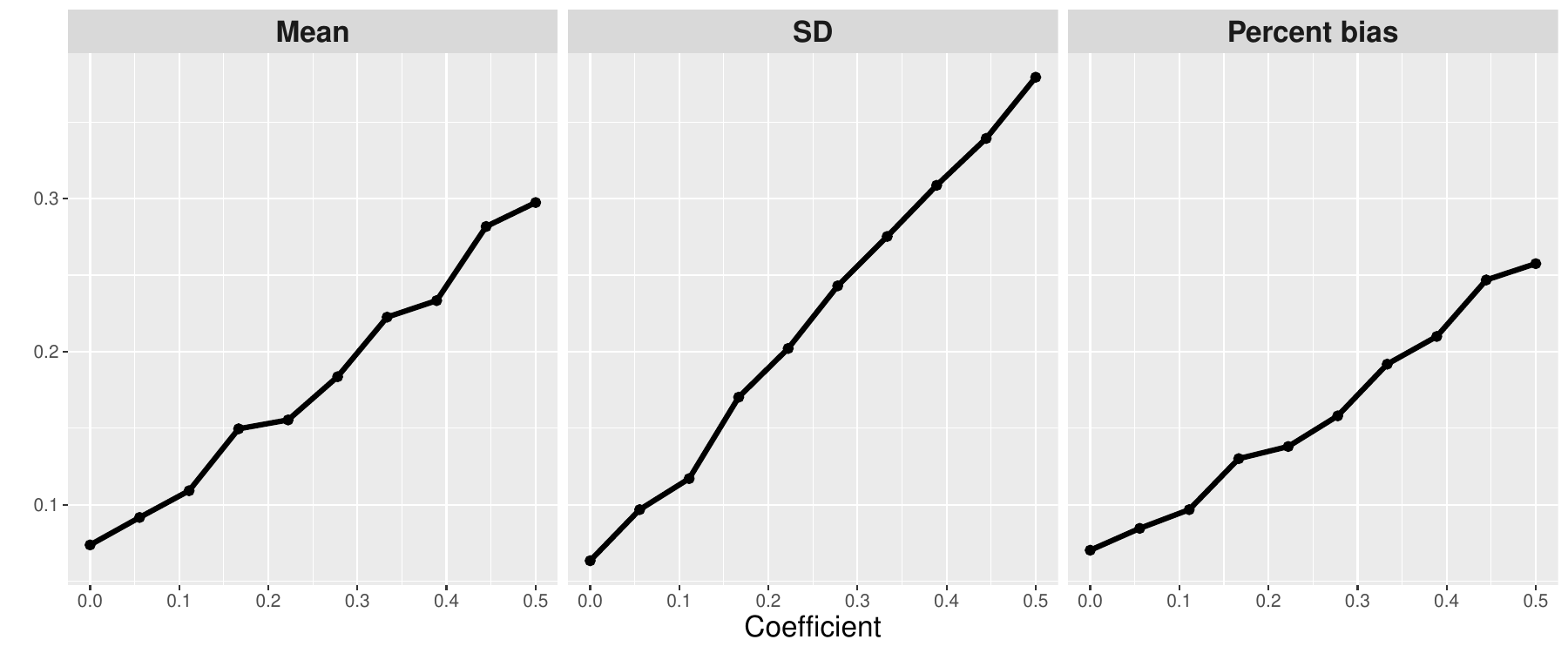} 
    \caption{Values of the sensitivity parameters under varying degrees of violations of identifying assumptions in the simulation scenarios of Section \ref{ssec:Interpretation}. The top panel shows the results under violations of Assumption \ref{ass:RacePlaceIgnore}, while the bottom panel shows the results under violations of Assumption \ref{ass:RacePlaceRatio}. Note that the mean value corresponds to how much the mean sensitivity parameter deviates from 1.}
    \label{fig:InterpretabilityMY}
\end{figure*}

The results can be found in the top panel of Figure \ref{fig:InterpretabilityMY}, and there are a couple of important takeaways from this analysis. As expected, we see larger values of the sensitivity parameters based on $\widetilde{\xi}(r, z)$ (referred to as the benchmark estimand) when compared with those found from $\xi(r, x)$. This provides empirical justification for benchmarking based on $\widetilde{\xi}(r, z)$, though also shows that conservative bounds will be obtained when using these benchmark values. This analysis also reveals that our primary estimand is relatively robust to unmeasured confounding bias. The largest coefficient value of 1 corresponds to very large logistic regression coefficients, yet the bias of our estimand is still relatively small at this degree of confounding. Moderate coefficient sizes of 0.5 lead to very small amounts of bias for the estimand of interest. Overall, this analysis shows that large values of the sensitivity parameters will only be obtained under very high degrees of association between $U$ and the observed variables. 

\subsubsection{Violations of Assumption \ref{ass:RacePlaceRatio}}

In this section, we explore violations of Assumption \ref{ass:RacePlaceRatio}. Unlike in the previous section, these assumptions need not be violated by the presence of an unmeasured variable $U$, and can be violated due to differential impacts of the covariates on $R$ in the $M=1$ population compared with the full population. We again explore a simple simulation study where all of the observed variables follow linear or logistic regression models. Given that violations of this assumption occur due to the effects of the covariates varying across different subset of the population, we explore violations by adding interaction terms between $R$ and $X$ in the models for generating $M$ and $Y$. We again use the same value of this constant for each interaction term and then vary its magnitude between 0 and 0.5. We look at a smaller interval here up to 0.5 as interaction terms are generally expected to be smaller in magnitude than main effect terms, and this still corresponds to rather large coefficients for interaction terms in a logistic regression model. Benchmarking with observed covariates is only applicable to violations of Assumption \ref{ass:RacePlaceIgnore}, and therefore we only present results for our primary estimand in this setting. 

The results can be found in the bottom panel of Figure \ref{fig:InterpretabilityMY}. The first thing to note is that our estimand appears to be less robust to violations of Assumption \ref{ass:RacePlaceRatio} than Assumption \ref{ass:RacePlaceIgnore}. While it is difficult to compare coefficient magnitudes between the two scenarios, it is clear that non-negligible amounts of bias can occur when this assumption is violated. This analysis additionally helps to provide intuition to researchers looking to reason about potential magnitudes of the sensitivity parameters when considering violations of Assumption \ref{ass:RacePlaceRatio}. For example, this makes clear that the mean and standard deviation of $\xi(r, x)$ likely don't exceed 0.3 unless there are very strong violations of Assumption \ref{ass:RacePlaceRatio}.

\subsection{Upper bounds assuming monotonicity}
\label{ssec:LowerBounds}

As an alternative to the sensitivity analyses described above, one can also construct bounds for $\Psi(r,x)$ under a weakened set of assumptions.  One of the critical pieces for $\Psi(r,x)$ is Assumption \ref{ass:RacePlaceRatio}, which allows us to target probabilities in the general population that are conditional on $X$. If one is not willing to make this assumption, we can still provide bounds on causal effects under mild monotonicity assumptions. To this end, one can assume a form of monotonicity with respect to $r$ that states $P(M(1) = 1 \mid R=r, X=x)$ is a monotonically decreasing function of $r$. This is a mild assumption given the existing literature showing that minorities are more likely to be stopped in predominantly white areas \citep{rojek2012policing, hannon2021out, simes2023policing}. We then provide bounds for $\Psi(r,x)$ as follows:
\begin{align*}
    \Psi(r, x) &\geq \Psi_{\text{na\"ive}}(r, x), \quad \text{if } r \leq r_0 \\
    \Psi(r, x) &\leq \Psi_{\text{na\"ive}}(r, x), \quad \text{if } r \geq r_0, 
\end{align*}
where we define
$$\Psi_{\text{na\"ive}}(r, x) = \frac{E(Y \mid R=r, M=1, D=1, X=x)}{E(Y \mid R=r_0, M=1, D=1, X=x)}.$$
A proof of this result can be found in Appendix C. This na\"ive form of the estimand essentially captures the effect of race and place policing, but only once a Black individual has already been stopped by the police. It precludes the possibility that Black individuals are more likely to be stopped in predominantly white areas, which may be a significant driver of any race and place effect. Importantly, these bounds only rely on Assumption \ref{ass:RacePlaceIgnore} holding, and do not rely on Assumption \ref{ass:RacePlaceRatio}. Additionally, there are no secondary data sources required to estimate these bounds as they only depend on the data for which $M=1$. Overall, these bounds show the following: If race and place policing is present in the sense that Black individuals are more likely to have force used against them in predominantly white areas, then the true value of $\Psi(r,x)$ is decreasing in $r$. In this case, the na\"ive estimand will either underestimate the magnitude of this effect returning functions that are flatter, or it will show an effect in the reverse direction with estimated curves that are increasing in $r$. Therefore, if the na\"ive estimand shows a decreasing function of $r$, then race and place policing can be identified, even if Assumption \ref{ass:RacePlaceRatio} does not hold.

\section{Simulation studies}

Here we conduct simulation studies to empirically investigate the properties of our proposed procedure and to evaluate how it performs under violations due to unmeasured confounding. In this section, we will estimate both CRR$(x)$ as well as the race and place estimand given by $\Psi(r,x)$. 

\subsection{Simulation design}
We simulate data under two different scenarios: one where all identification assumptions are met, and one where assumptions are violated due to the presence of an unmeasured confounder. In each data set there are $q=35$ precincts with a total sample size of $n=100000$. We set $q_i$ to be a numeric value indicating which precinct observation $i$ falls in, where each observation is randomly assigned to one of the 35 precincts. We first generate a single categorical covariate $X$ which takes values 1 to 4 with equal probability. In the first simulation scenario, we generate values for $D_i$, $M_i(d)$, and $Y_i(d)$ from Bernoulli distributions with probabilities given by
\begin{align*}
    P(D_i=1 \mid X=x) &= \text{Sigmoid}(2q_i/q-1)  \\
    P(M_i(d) = 1 \mid X=x) &= \text{Sigmoid}(x-2.5) \text{ for } d=0,1 \quad  \\
    P(Y_i(d,1) = 1 \mid X=x, R=r) &= \text{Sigmoid}(x-0.5- 2r) \text{ for } d=0,1. \\
    \text{where} \quad P(q_i = j)  &= \frac{1}{q}, \quad j = 1, 2, 3, \ldots, q.
\end{align*}
Note that $R_i$ is subsequently defined as the proportion of the data points with $D_i = 1$ in the precinct that observation $i$ falls in. Further, we use the logistic sigmoid given by $\text{Sigmoid}(a) = \frac{1}{1 + e^{-a}}$, so that the binary variables are generated from logistic regression models. 

In order to assess the impact of violations to key assumptions, we also generate data under a second scenario with an unmeasured confounder that affects both race and whether they are likely to have force used against them. One can think of this unmeasured confounder as representing criminologic behavior, which may be higher in a particular race, but also make it more likely for force to be applied against them. The data generating process for this simulation is given by
\begin{align*}
    &U_i \sim \text{Unif}(0,1) \\
    &P(D_i=1 \mid X=x, U=u) =\\& \quad \text{Sigmoid}(u+2*q_i/q-1) \\
    &P(M_i(d) = 1 \mid X=x, U=u)=\\& \quad \text{Sigmoid}(x-2.5) \text{ for } d=0,1 \\
    &P(Y_i(d,1) = 1 \mid X=x, U=u, R=r)=\\& \quad u*\text{Sigmoid}(x-0.5- 2*r)  \text{ for } d=0,1.
\end{align*}
Note that in both simulations the true causal risk ratio is 1 for all values of $X$, while the race and place estimand varies across different $r$ values.

Throughout we use a specific choice of $x$ to evaluate both estimands, but similar results are found for other values of the covariates. To evaluate performance across a range of settings, we calculate both percent bias and mean squared error (MSE) averaged over 100 simulated data sets. For $\Psi(r,x)$ we average these metrics across a grid of $r$ values that spans the domain of $R$ in our simulation studies, which here is between 0.2 and 0.8. Lastly, we need to specify a variety of models for each of the two estimators. For all models with binary outcomes, we utilize standard logistic regression models. In the race and place estimand, we also require modeling the density of $R$, with R being a continuous number on the unit interval. We explore both a kernel density estimator using a t-distribution kernel with 2 degrees of freedom, as well as an estimator based on beta regression.  We only display results for the kernel density estimator here, but the results from beta regression can be found in Appendix B. Overall, beta regression provides similar results, though it performs worse in the tails of $R$, and therefore we focus on the more flexible kernel density estimator here. 

\begin{figure*}[!t]
\centering
    \includegraphics[scale=0.47]{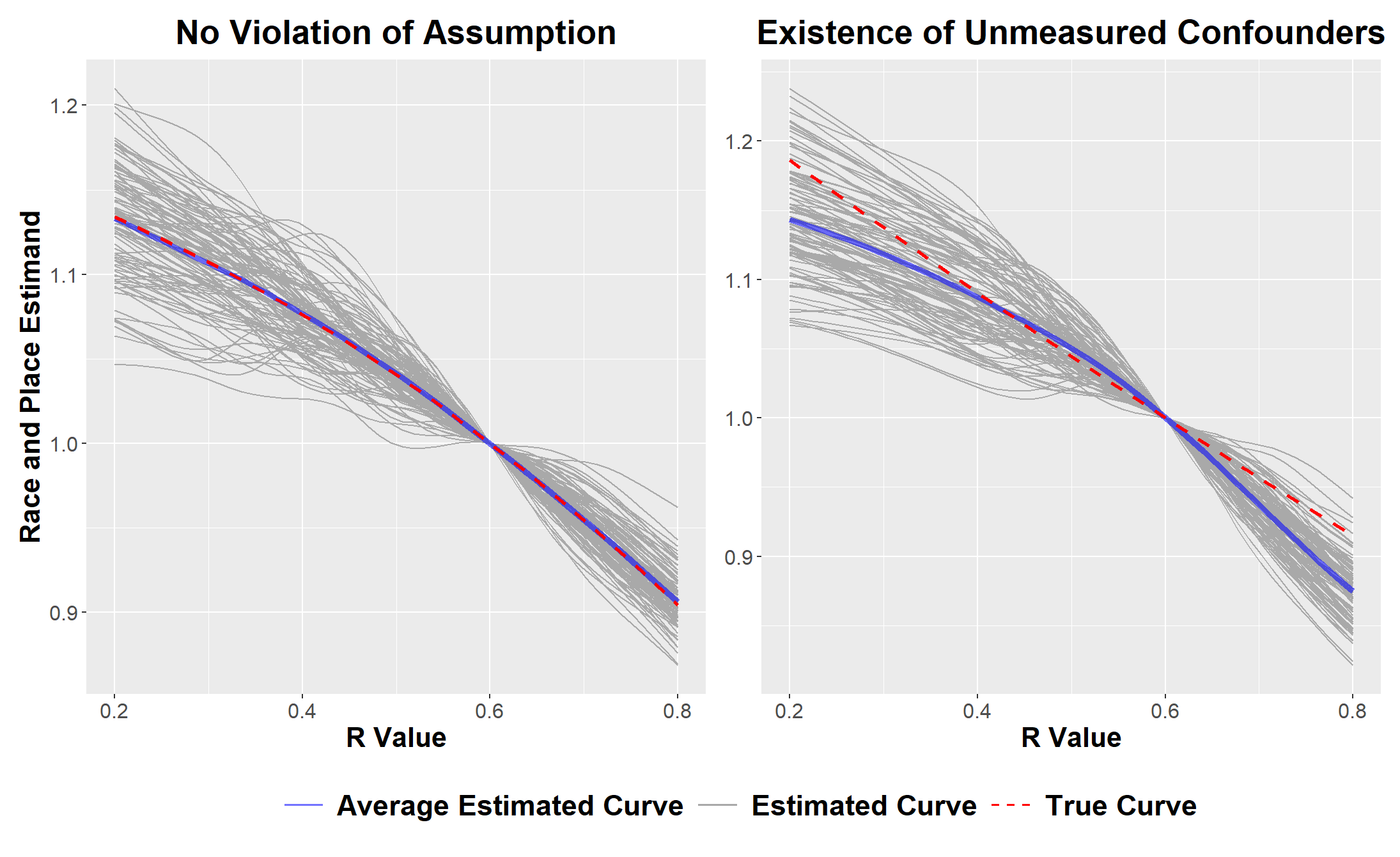} 
    \caption{The left plot contains estimates of $\Psi(r,x)$ across simulated data sets when there are no unmeasured confounders. The right plot shows estimates of $\Psi(r,x)$ across simulated data sets when there are violations of Assumption \ref{ass:RacePlaceIgnore}, with an unmeasured confounder affecting both race and whether police use force against someone. Individual grey lines correspond to estimates from a single data set, while the blue line is the average across 100 repetitions of the same simulation.}
    \label{fig:SimulationKDE}
\end{figure*} 
\begin{figure}[h]
    \includegraphics[width=0.95\linewidth]{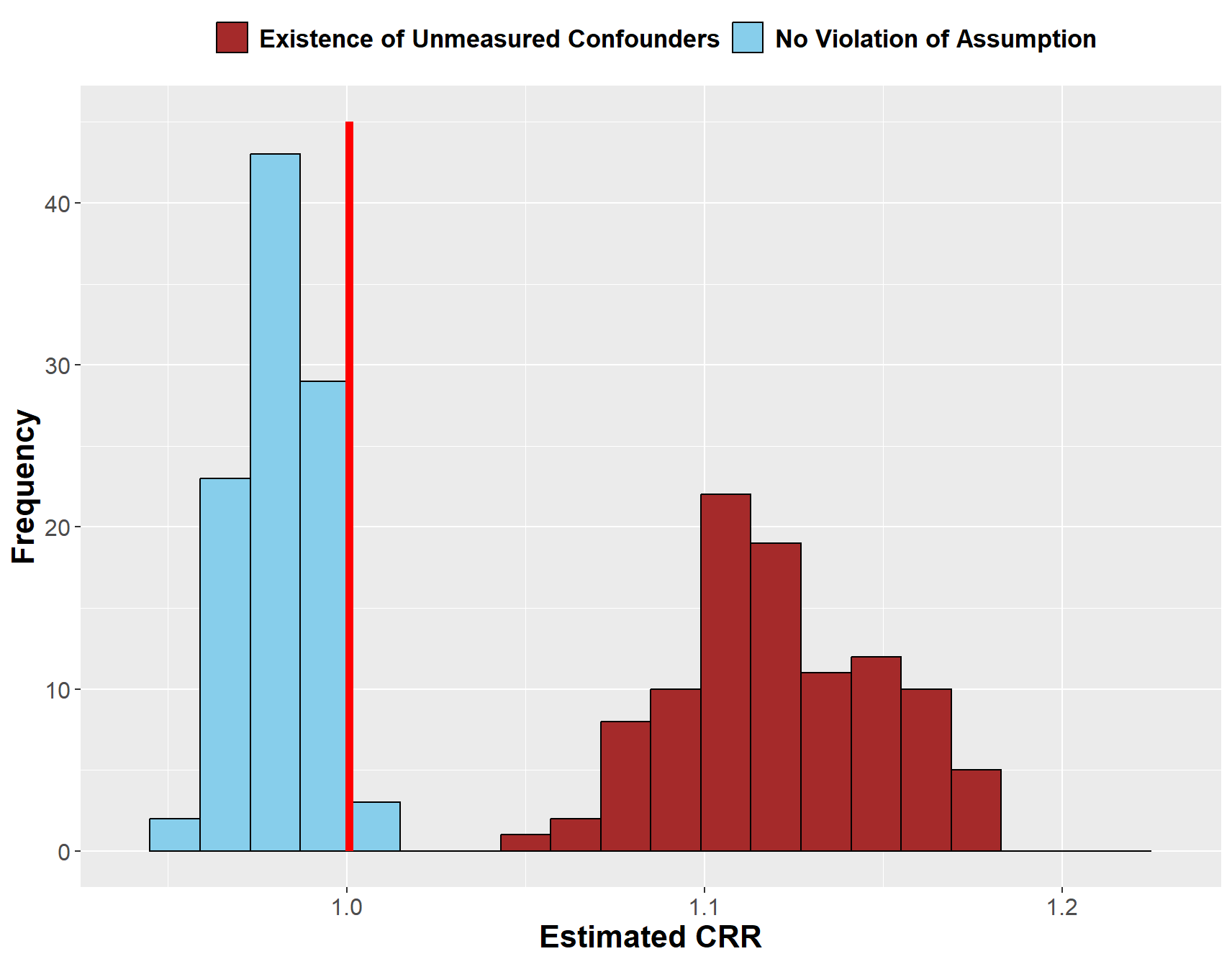} 
    \caption{Distribution of CRR$(x)$ estimates across all simulated data sets. The blue histogram corresponds to the situation with no unmeasured confounders, while the red histogram is from the simulation with unmeasured confounding. The true CRR is given by the vertical line. }
    \label{fig:SimulationCRR}
\end{figure} 

\subsection{Results} 

The results averaged across all 100 simulated data sets can be found in Figure \ref{fig:SimulationKDE}, Figure \ref{fig:SimulationCRR}, and Table \ref{tab:SimulationResults}, and there are a few main points to discuss. As expected, when there are no unmeasured confounders, estimation for both the causal risk ratio and the race and place estimand is nearly unbiased with estimates close to the true value. While this result is expected, it confirms our estimation strategy for both estimands is reasonable and shows that the flexible kernel density estimator is capturing the conditional density of $R$ reasonably well. Of more interest is what happens in the second scenario with an unmeasured confounder $U$ that violates Assumption \ref{ass:RacePlaceIgnore} as well as the corresponding assumption for the CRR. We see bias in both of the estimands, but there are considerable differences between the causal risk ratio and the race and place estimand. CRR$(x)$ has a fairly large percent bias, as illustrated in both Figure \ref{fig:SimulationCRR} and Table \ref{tab:SimulationResults}, and importantly this leads to anti-conservative estimates of CRR$(x)$ that are all above 1, despite the true value being the null value of 1. The race and place estimand, $\Psi(r,x)$ shows bias for certain regions of $r$, but this percent bias is quite low. This highlights the somewhat weakened identification conditions required for this estimand, and its robustness against unmeasured confounding. It should be clarified that this is not a direct comparison between the CRR and $\Psi(r,x)$ or a criticism of the CRR. They are simply two fundamentally different estimands that target different aspects of racial bias, and this simulation highlights that $\Psi(r,x)$ might be more robust to certain types of unmeasured confounding.

\begin{table}[h]
\begin{tabular}{||c c c||} 
\hline
 \multicolumn{3}{|c|}{No violation of assumptions}\\ 
 \hline
 &Average \% Bias & MSE \\ 
 \hline
 $\Psi(r,x)$ & 1.26\% & 0.0003658978\\
 \hline
 CRR$(x)$ & 1.93\% & 0.0004819435\\
 \hline
 \multicolumn{3}{|c|}{Unmeasured confounder leading to violations of assumptions}\\ 
 \hline
 $\Psi(r,x)$ & 2.21\% & 0.0009701268\\

 \hline
 CRR$(x)$ & 11.65\% & 0.01430926\\
 \hline
\end{tabular}
\caption{Average percent bias and MSE for both estimands in both simulation scenarios.}
\label{tab:SimulationResults}
\end{table}

\section{Analysis of NYPD stop data}
\label{sec:NYPDanalysis}

In this section we re-analyze the NYPD ``Stop-and-Frisk" data set provided in \cite{fryer2019empirical} and further analyzed in \cite{zhao2022note}. The NYPD Stop and Frisk Data records police stops, with detailed information on both the officer and suspect. With over 5 million entries over the time period 2003-2013, this data provides detailed records for research on policing disparities, although importantly only restricts to police encounters that led to a stop ($M=1$). We dichotomize race throughout so that $D_i=1$ corresponds to a Black individual, while $D_i=0$ corresponds to a white individual, and we drop stops of individuals of other races from our data set. In total the data set contains 18 covariates providing contextual information about the stop as well as certain characteristics of the individual being stopped by the police. These covariates include the time of day, location in the city, and a number of other features describing the type of stop and the individual being stopped. Due to the larger number of covariates, we do not use the kernel density estimators used in the simulation, and focus on conditional density estimators based on beta regression for estimating any conditional densities for $R$. Further, we estimate both the causal risk ratio and $\overline{\Psi}(r)$ here. Our first goal is to inspect the impact of incorporating anonymized and aggregated cell phone mobility data into estimation of CRR$(x)$ to see if racial demographics of the city change drastically due to mobility and whether this leads to different estimates of racial bias in policing. We then evaluate a different question when examining $\overline{\Psi}(r)$, which is whether Black individuals are policed differently in predominantly white neighborhoods compared with predominantly Black neighborhoods. All confidence intervals are 95\% confidence intervals and are obtained using the nonparametric bootstrap.   

\subsection{Potential shortcomings of mobility data}

Before discussing the results of our analyses, it is first important to acknowledge potential limitations of using mobility data in a study such as this, and how these limitations might impact our results. We incorporate anonymized and aggregated cell phone mobility data in our study, though similar issues would hold for any source of population level mobility information. One potential issue is the representativeness of mobility data given differential amounts of cell phone ownership across demographic groups, but recent studies have found cell phone mobility data to be largely representative of the overall population \citep{sinclair2023assessing}. Additionally, within NYC specifically, a recent study in 2015 found that 79\% of all residents owned a smart phone \citep{de2015new}. This provides some evidence that our results won't be drastically impacted by a lack of representativeness of mobility data with respect to the population being studied. 

A potentially bigger concern is that our study covers data from 2003 to 2013, while the mobility data used comes from a more recent time period of 2019. This could bias our results if mobility patterns within NYC have changed drastically between these time periods, though there is some evidence to suggest this is not a major issue. For one, commuting methods in the city were stable during this time period. In 2000, the average commute time in the city was 40 minutes and 53\% of residents commuted by public transit. In 2019, the figures were 42 minutes and 57\% \citep{acs2020}. The spatial arrangement of city demographics was also stable during this time. Between 2000 and 2017, racial segregation in the U.S. saw only ``modest reductions of 1 to 4 points" \citep{frey2018blackwhite_segregation}. Even gentrification, among the most high-profile types of urban restructuring, was confined to a relatively few neighborhoods. Only 6\% of census tracts in New York City gentrified between 2000 and 2015 \citep{maciag2015gentrification}. Lastly, we incorporate a secondary source of mobility data, which is available over the entire study period, to examine the extent to which mobility patterns have changed over this time period. Specifically, we examined the United States Census LODES data (Longitudinal Employer-Household Dynamics Origin-Destination Employment Statistics, \citep{lehd_data}), which provides information at the Census tract level about where individuals both live and work. This allows us to study commuting patterns over time for individuals who commute to work, which will highlight the extent to which mobility within NYC has changed over time.  This analysis can be found in Appendix J, where we find that there are small changes in the amount and direction of mobility, but the prevailing commuting patterns are similar between 2003 and 2019. The stability of New York City’s commuting patterns and demographic geography reassures us that more recent mobility data still adequately reflects earlier time periods. It is also likely that these estimates of mobility are more reflective of mobility patterns in NYC than simply ignoring mobility altogether, which is common practice in related studies \citep{zhao2022note}. 

Despite the aforementioned justification for using this data source, this is still a potential drawback of the study, and therefore we additionally assess sensitivity to misspecified mobility patterns. Specifically, if our mobility data is not representative of the population of interest, then Assumption \ref{ass:RacePlaceRatio} will be violated, and we assess robustness to this assumption in Section \ref{ssec:NYPDsens3}. As a separate check of robustness, we also estimate $\overline{\Psi}(r)$ using only data from 2012 and 2013, which represents a more similar time frame as the mobility data. These results can be found in Appendix I, where we find similar conclusions as those found on the full data. 

\subsection{Estimating the CRR using mobility data}

\begin{figure*}[htbp]
\centering
    \includegraphics[scale=0.6]{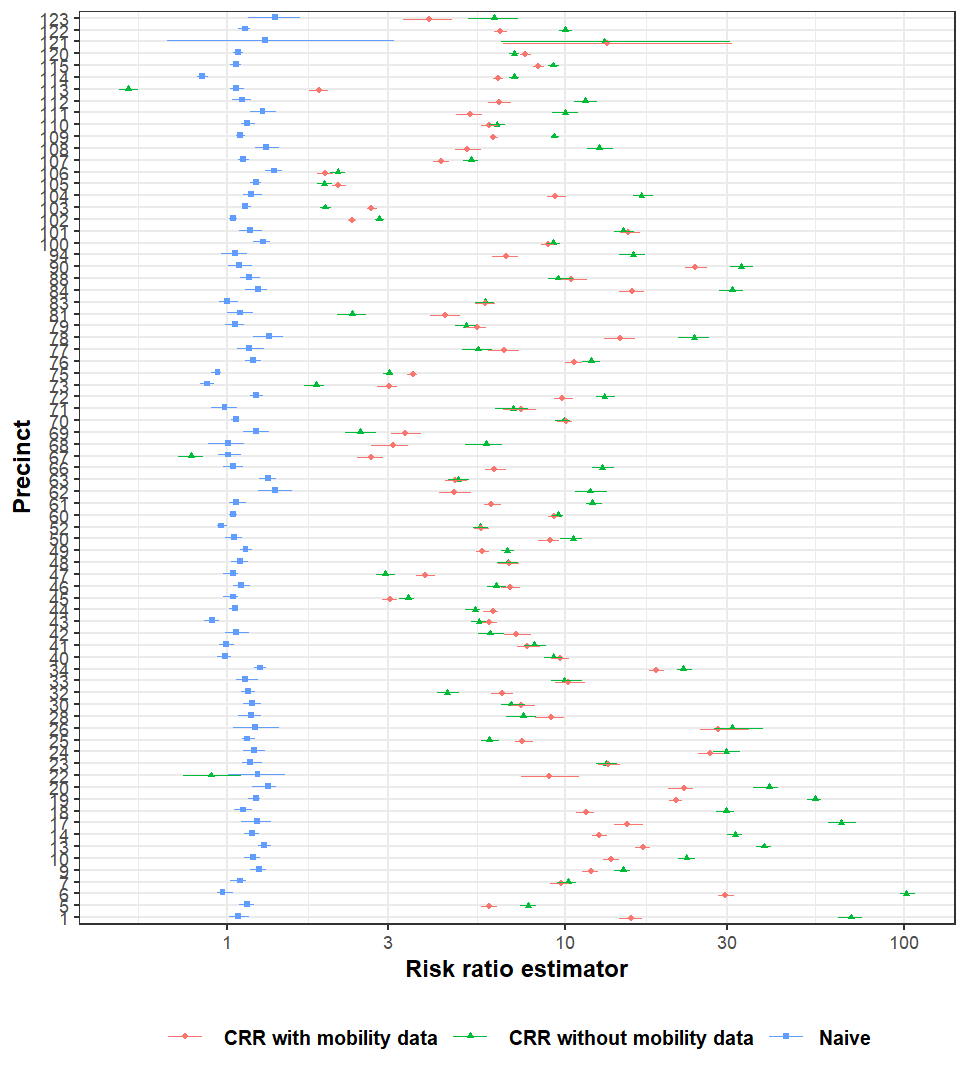} 
    \caption{Estimates and  95\% confidence intervals for all estimators of the CRR separated by precinct. The na\"ive estimator is described above and only accounts for racial bias once an individual has been stopped. The other two estimators are the CRR estimated with and without mobility data to better understand the population encountering the police in each precinct.}
    \label{fig:CRRall}
\end{figure*}

We first estimate CRR$(x)$, which was first defined in \cite{zhao2022note} though there are some key differences here. For one, we incorporate anonymized and aggregated cell phone mobility data, which should provide an improved estimate of $P(D = d | X = x)$, a key component when estimating CRR($x$). Second, we averaged over the empirical distribution of the covariates to obtain an overall causal risk ratio, which we define as CRR $= 1/n \sum_{i=1}^n \text{CRR}(X_i)$. Note that we estimate a separate CRR per precinct, which is shown in Figure \ref{fig:CRRall}. The results include na\"ive estimates, which average the following quantity over the empirical distribution of $X$ within each precinct:
$$\frac{E(Y \mid M=1, D=1, X=x)}{E(Y \mid M=1, D=0, X=x)}.$$
This term represents a more local effect that only captures the degree of racial bias once an individual has been stopped, and ignores any bias stemming from the likelihood of being stopped. We see these values are typically greater than 1, though are much closer to 1 than the estimators which aim to adjust for bias in the probability of being stopped. There is also a large difference in the CRR estimates that incorporate mobility versus those that do not, indicating that accounting for population mobility is a crucial aspect of racial bias in policing research. 
\begin{figure*}[htbp]
\centering
    \includegraphics[scale=0.44]{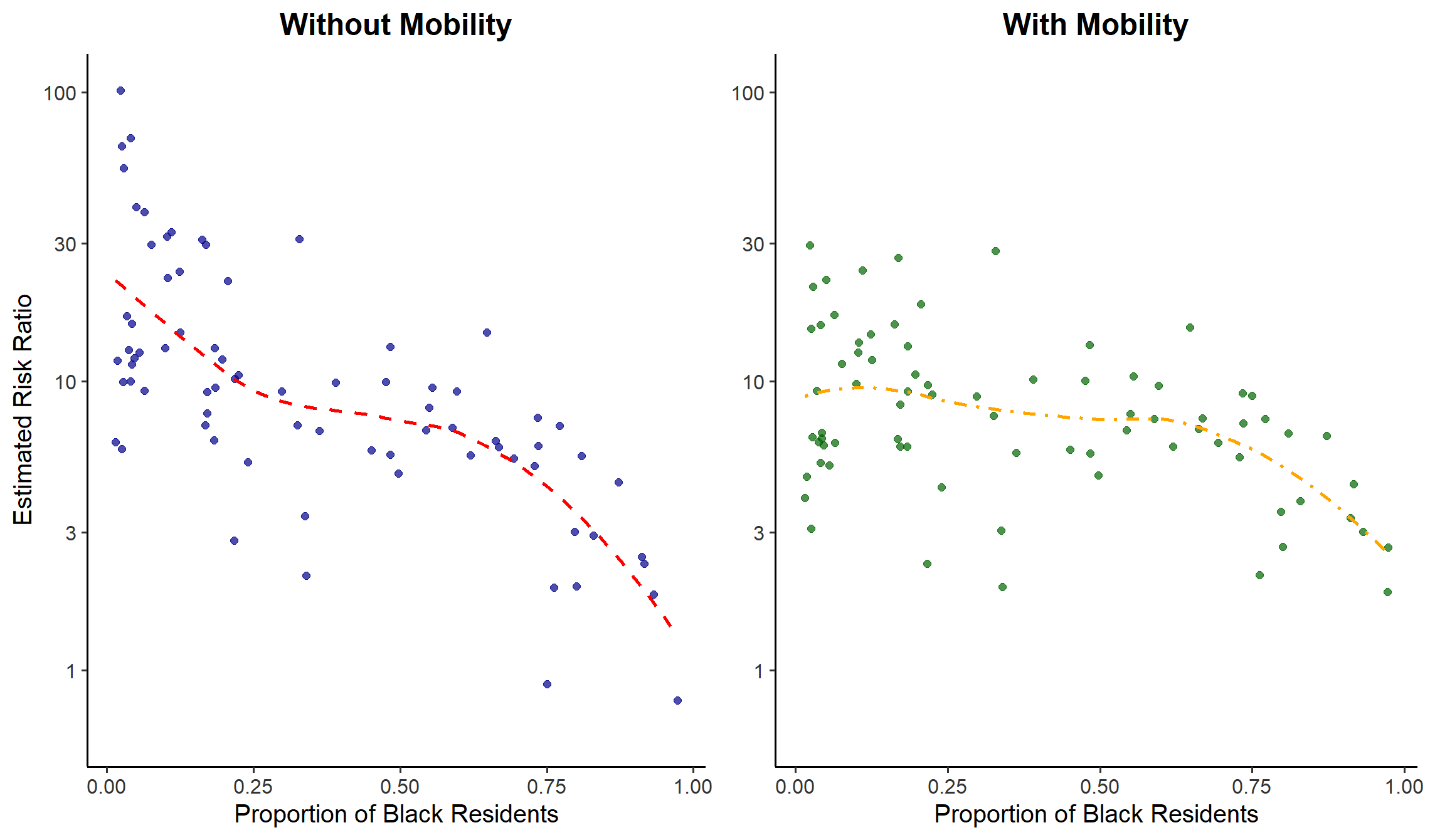} 
    \caption{Precinct-level CRR estimates as a function of the proportion of Black residents in the precinct. The left panel estimates the CRR without mobility data, while the right panel incorporates anonymized and aggregated cell phone mobility data.}
    \label{fig:CRRmob}
\end{figure*} 
We are also able to see this difference from incorporating mobility when looking at Figure \ref{fig:CRRmob}, which shows the precinct level estimates of the CRR as a function of the percentage of Black individuals within a precinct. We see that ignoring mobility leads to the conclusion that there is significantly less racial bias (in terms of the CRR) in areas of the city with larger proportions of Black individuals. Once mobility is accounted for, as seen in the right panel of Figure \ref{fig:CRRmob}, this conclusion weakens dramatically as the CRR does not vary greatly by the proportion of Black individuals in a precinct, though it does still does appear that areas with very high ($> 0.75$) proportions of Black individuals have smaller CRR estimates. These results point to the possibility that individuals are policed differently in different areas of the city, but do not directly tackle the question of race and place policing that focuses only on whether individuals of a particular race are policed differently across the city. For this, we examine $\overline{\Psi}(r)$ in the following section to examine this possibility.  

\subsection{Evaluating race and place}

\begin{figure}[h]
    \includegraphics[width=0.95\linewidth]{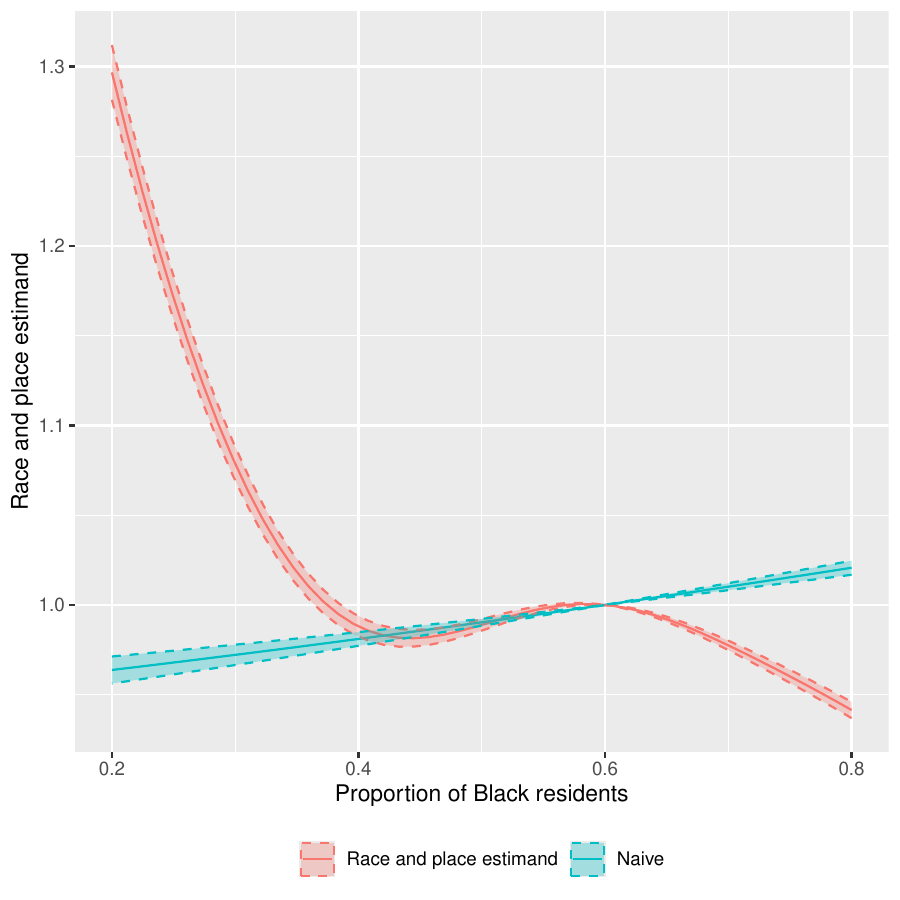} 
    \caption{Estimates and 95\% confidence intervals for $\overline{\Psi}(r)$ and the na\"ive estimator of this quantity obtained by averaging $\Psi_{\text{na\"ive}}(r, x)$ over the distribution of $X$.}
    \label{fig:RAP}
\end{figure}

Figure \ref{fig:RAP} presents the estimates and corresponding 95\% confidence intervals for $\overline{\Psi}(r)$, where we have set $r_0 = 0.6$, though this choice does not impact the overall findings. We can see that $\overline{\Psi}(r)$ is generally a decreasing function of $r$, which indicates that Black individuals are more likely to have force used against them if they are in areas with larger proportions of white residents, which suggests the existence of race and place policing. This is a rather large difference as well, as Black individuals are 1.38 times as likely to have force used against them when they are in a precinct with 20\% Black residents compared with when they are in a precinct with 80\% Black residents. Figure \ref{fig:RAP} also shows a na\"ive estimator, which is $\Psi_{\text{na\"ive}}(r, x)$ after averaging over the covariate distribution. As discussed in Section \ref{ssec:LowerBounds}, this na\"ive curve is a lower bound for $\Psi(r, x)$ when $r \leq r_0$, and an upper bound for $\Psi(r, x)$ when $r \geq r_0$. We see that this na\"ive estimate is substantially different from the estimate of $\overline{\Psi}(r)$, and even suggests an effect in the opposite direction, which is that Black individuals are less likely to have force used against them in predominantly white areas. This shows how critically important it is to account for selection bias when estimating racial bias from administrative data. Once selection bias is accounted for, the sign of the effect changes, and we see large effects suggesting that Black individuals are more likely to have force used against them in predominantly white neighborhoods. This also highlights that race and place policing occurs more in the decision to stop a Black individual, and less in the decision of whether to subsequently use force against them. Our analysis indicates that Black individuals are more likely to be stopped in predominantly white areas, but using the na\"ive estimate ignores this effect and leads to different conclusions.

\subsubsection{Sensitivity to unmeasured confounding using benchmarking}

\begin{figure*}[htbp]
\centering
    \includegraphics[width=0.85\linewidth]{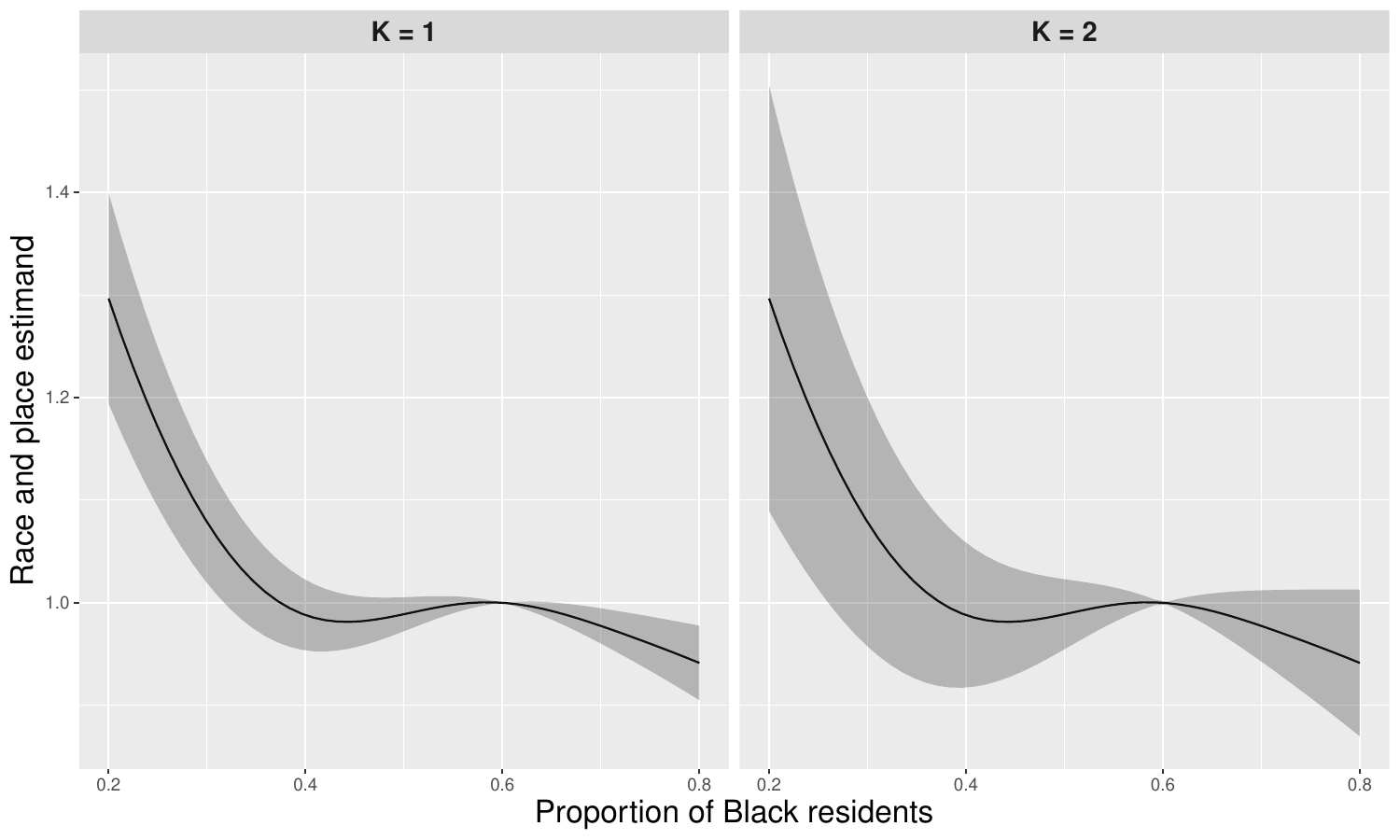} 
    \caption{Upper and lower bounds for $\overline{\Psi}(r)$ accounting for violations of Assumption \ref{ass:RacePlaceIgnore} when using benchmarking. The left plot shows the results when we assume the unmeasured variable has the same strength of associations as the strongest observed covariate, and the right plot shows the results when it has twice the strength of association as the strongest observed covariate.}
    \label{fig:Sensitivity1}
\end{figure*}

We now use the formal benchmarking procedure described in Section \ref{ssec:Benchmark} to evaluate the robustness of our results to unmeasured confounding and violations of Assumption \ref{ass:RacePlaceIgnore}. We perform the proposed benchmarking procedure with each of our covariates, and we retain the results from the covariate with the largest values of the sensitivity parameters, which in this case corresponds to the binary variable of whether police are in uniform. Throughout, we assume the correlation sensitivity parameter is fixed to 1 or -1, whichever leads to the worst bias. We evaluate the sensitivity of our results when we assume 1) the unobserved covariate has the same strength of association with the observed variables as the observed covariate ($K = 1$), and 2) it has twice the strength of association as the observed covariate ($K = 2$). These results can be found in Figure \ref{fig:Sensitivity1}, which show bounds for $\overline{\Psi}(r)$ under these two assumptions on the strength of the unmeasured confounder. We see that an unmeasured confounder with the same strength of associations as the strongest observed confounder does not bias the estimated causal effect substantially. Even when we allow the unmeasured confounder to have twice the strength of association as the strongest observed confounder, the results are robust in the sense that there is still a negative trend in the function when going from $r=0.2$ to $r=0.8$, which suggests the existence of race and place policing. These results show that only a very strong unmeasured confounder could entirely explain our findings, and it would have to be significantly stronger than any of the observed covariates within our data set. As a reminder, our benchmarking is based on an estimand that typically produces quite conservative benchmarking values for our estimand of interest, and therefore the degree of confounding required to eliminate our estimated effect is likely even stronger than this suggests. 

\subsubsection{Sensitivity to Assumption \ref{ass:RacePlaceRatio}}
\label{ssec:NYPDsens3}

\begin{figure}[htbp]
\centering
    \includegraphics[width=0.99\linewidth]{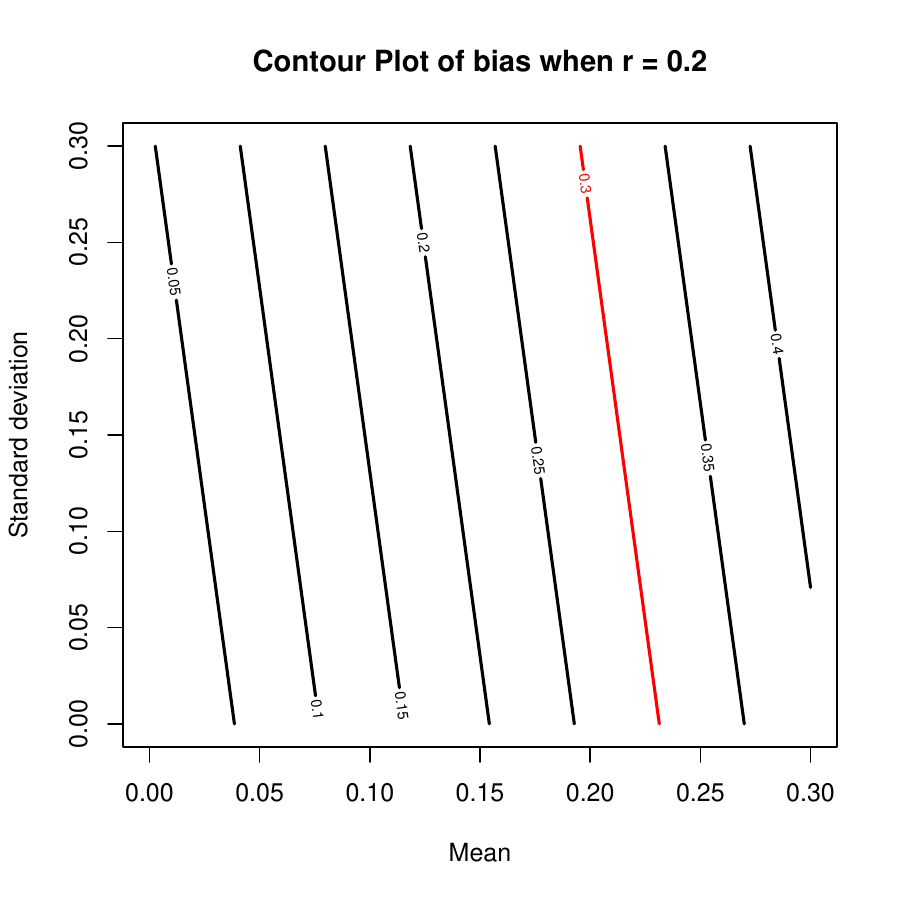} 
    \caption{Contour plot showing the magnitude of the maximum bias of $\overline{\Psi}(0.2)$ as a function of the two sensitivity parameters. The red line shows the contour at the estimated value of $\overline{\Psi}(0.2)$, highlighting how much bias is required to reduce the effect to the null value of 1. Note that the mean value corresponds to how much the mean sensitivity parameter deviates from 1. }
    \label{fig:Sensitivity3}
\end{figure}

While the benchmarking results of the previous section were useful for understanding the robustness to unmeasured confounding, they do not apply to evaluating robustness to Assumption \ref{ass:RacePlaceRatio}. This assumption can be violated for reasons not related to the omission of a confounder, such as 1) differential effects of the covariates on $R$ in the $M=1$ population versus the overall population, or 2) misspecification of population level mobility due to a lack of representativeness of the mobility data for our population of interest. To assess robustness to this assumption, we can examine how results would change under a multitude of sensitivity parameters, and then use the findings of our interpretability simulations in Section \ref{ssec:Interpretation} to understand whether these parameters correspond to large violations of this assumption. For this analysis, we focus on a particular value of $r$, which in this case we set to $r = 0.2$, the smallest value considered. In Figure \ref{fig:Sensitivity3}, we show contours of the maximum possible bias for $\overline{\Psi}(0.2)$ as a function of the two sensitivity parameters, again assuming the correlation sensitivity parameter is 1 or -1. The red line on the contour plot highlights the amount of confounding that would reduce our estimated effect of 1.3 down to the null value of 1. We can see that in this case, the bias is largely driven by how much the mean parameter $E(\xi(r, x))$ deviates from 1. Within the ranges of the sensitivity parameters explored in the figure, the mean sensitivity parameter must deviate from 1 by at least 0.2 in order to bias the estimated effect to the null value. Using Figure \ref{fig:InterpretabilityMY} as a way to help reason about the magnitudes of the sensitivity parameters when Assumption \ref{ass:RacePlaceRatio} is violated, this suggests a moderate degree of robustness to this assumption. Our results do not appear as robust to Assumption \ref{ass:RacePlaceRatio} as they are for Assumption \ref{ass:RacePlaceIgnore}, though it would still take fairly sizeable misspecification of the mobility data or large differences in covariate effects in the $M=1$ population in order to completely remove the estimated effect. 

\section{Conclusion}

Studying racial bias in policing is a crucially important problem, but one that comes with a number of inherent difficulties, particularly due to the nature of the observed data, which is limited to only those police-civilian interactions that led to a stop. We have contributed to this growing discussion in two distinct ways. First, we develop a novel estimand, which we refer to as the race and place estimand, that targets a fundamentally different question than most studies of racial bias in policing. This estimand examines whether police treat civilians of a particular race differently depending on the demographics of the neighborhood that the encounter occurs in. We show how to identify this estimand using observed data, and provide a sensitivity analysis framework to assess how robust our conclusions are to violations of those key, untestable assumptions. A second key contribution, is the incorporation of anonymized and aggregated cell phone mobility data, which provided us with improved estimates of the racial demographics of the city. This is a crucial component to any analysis of racial bias as it helps to better understand the racial distribution of people that are potentially encountering the police in any one location. We utilized this additional data source for the race and place estimand, but also showed that existing estimands, such as the causal risk ratio, have drastically different estimates once mobility is accounted for. This shows the importance moving forward of attempting to account for mobility of individuals across the city when studying racial bias in policing. 

As with any study of racial bias from administrative data, our study comes with inherent limitations. For either estimand considered, we relied on certain untestable assumptions. For one, we assume that there are no unmeasured confounders biasing our results. While this is alleviated somewhat since the data contained a significant amount of information about the stop and the civilian being stopped, this is still an unverifiable assumption. Additionally, both estimands relied on estimates of the racial demographics, \textit{conditional on covariates}, as given by $P(D=d \mid X=x)$. This is not estimable from the observed data, which only conditions on $M=1$, so we incorporated two secondary data sources: the United States Census Data, and anonymized and aggregated cell phone mobility data. These secondary data sources provide only marginal information about the demographics of the city, and we relied on an assumption to link these to the conditional distributions that we require. We were able to test the robustness to all of these assumptions through a sensitivity analysis that allowed for violations of these assumptions up to a certain magnitude, which provides additional evidence of a causal effect. Despite this, future research should focus on incorporating improved data sources that make less restrictive assumptions necessary to identify the causal effect of race on policing outcomes. Additionally, we focused on identification and sensitivity analysis throughout, but an area for future research would be to develop improved estimators for the estimands considered. We used plug-in estimators of all conditional expectations or densities required in our identification formula, but estimators based on semiparametric efficiency theory could provide increased robustness and efficiency.

\bibliographystyle{apalike}
\bibliography{References} 

\appendix

\section{Identification of $\Psi(r,x)$}

Now we can write the numerator of our estimand as
\begin{align*}
    E[Y(1) \mid R=r, X=x] &= E[Y(1) \mid R=r, X=x, M(1) = 1] \ P(M(1) = 1 \mid R=r, X=x) \\
    & \ + E[Y(1) \mid R=r, X=x, M(1) = 0] \ P(M(1) = 0 \mid R=r, X=x) \\
    &= E[Y(1,1) \mid R=r, X=x, M(1) = 1] \ P(M(1) = 1 \mid R=r, X=x) \\
    & \ + E[Y(1,0) \mid R=r, X=x, M(1) = 0] \ P(M(1) = 0 \mid R=r, X=x) \\
    &= E[Y(1,1) \mid R=r, X=x, M(1) = 1] \ P(M(1) = 1 \mid R=r, X=x) \\
    &= E[Y \mid R=r, X=x, M = 1, D=1] \ P(M(1) = 1 \mid R=r, X=x) c_1(x) \\
    &= E[Y \mid R=r, X=x, M = 1, D=1] \ P(M = 1 \mid R=r, X=x, D=1) c_1(x) c_2(x)
\end{align*}
Now focus on $P(M = 1 \mid R=r, X=x, D=1)$:

\begin{align*}
    P(M = 1 \mid R=r, X=x, D=1) = \frac{P(D = 1 \mid R=r, X=x, M=1) P(M=1 \mid R=r, X=x)}{P(D=1 \mid R=r, X=x)},
\end{align*}
which can be further decomposed as
\begin{align*}
    \frac{P(D = 1 \mid R=r, X=x, M=1) f(r \mid M=1, X=x) P(M=1 \mid X=x)}{P(D=1 \mid R=r, X=x) f(r \mid X=x)}.
\end{align*}
If we combine these results, and take the ratios of the values at $R=r$ and $R=r_0$, we see that
\begin{align*}
    \frac{E[Y(1) \mid R=r, X=x]}{E[Y(1) \mid R=r_0, X=x]}
    &= \frac{E(Y \mid R=r, M=1, D=1, X=x)}{E(Y \mid R=r_0, M=1, D=1, X=x)} \frac{P(D=1 \mid R=r, M=1, X=x)}{P(D=1 \mid R=r_0, M=1, X=x)} \\
    &\times \frac{f(r \mid M=1, X=x)}{f(r_0 \mid M=1, X=x)} \frac{P(D=1 \mid R=r_0, X=x) f(r_0 \mid X=x)}{P(D=1 \mid R=r, X=x) f(r \mid X=x)} \\
    &= \frac{E(Y \mid R=r, M=1, D=1, X=x)}{E(Y \mid R=r_0, M=1, D=1, X=x)} \frac{P(D=1 \mid R=r, M=1, X=x)}{P(D=1 \mid R=r_0, M=1, X=x)} \\
    &\times \frac{f(r \mid M=1, X=x)}{f(r_0 \mid M=1, X=x)} \frac{f(r_0 \mid D=1, X=x)}{f(r \mid D=1, X=x)}
\end{align*}
Note that crucially, the $c_1(x)$ and $c_2(x)$ terms cancel out in both the numerator and denominator and therefore this estimand is robust to violations of unconfoundedness as long as these violations are not a function of $R$.

\section{Simulation results using beta regression}

In this section, we utilize beta regression \citep{ferrari2004beta} to replace the role of kernel density estimation and estimate race and place estimands. Kernel density approaches are completely non-parametric, which allows for a high degree of flexibility. It can become increasingly inefficient, however, as the dimension of $X$ increases. Beta regression on the other hand scales relatively well to the addition of more covariates, as we have in our motivating application, so we study its performance in simulation here. To study the properties of our proposed procedure in using beta regression, we conduct simulation studies under the same settings as in the manuscript. We do not alter any logistic regression models built for models with binary outcomes. Instead, we focus solely on modeling the density of $R$ using beta regression. Overall, beta regression provides similar results, though it performs slightly worse in the tails of R.

 \begin{figure}[h]
 \centering
    \includegraphics[width=0.9\linewidth]{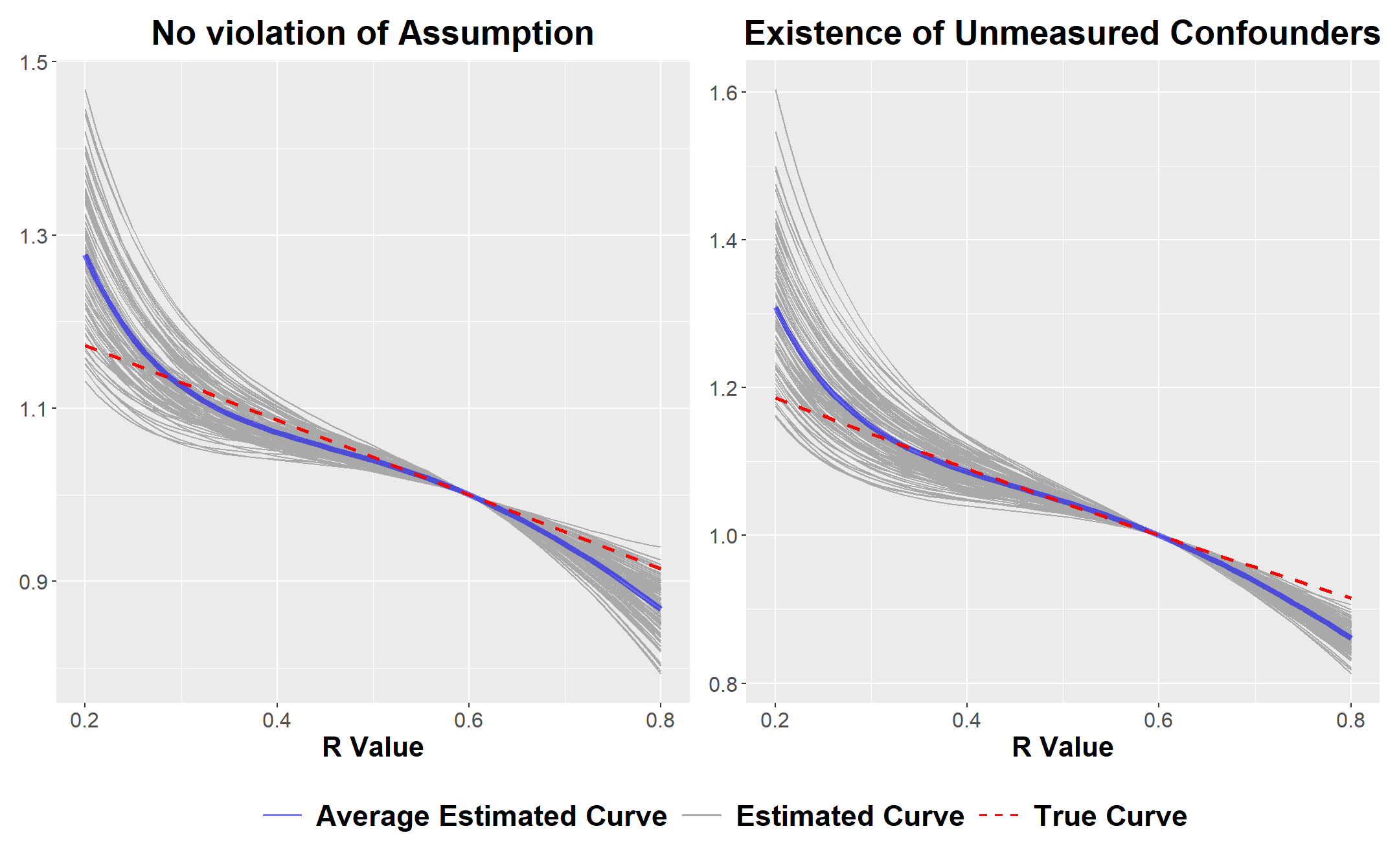} 
    \caption{Simulation result when using beta regression instead of kernel density estimation. The left plot contains estimates of $\Psi(r,x)$ across simulated data sets when there are no violations of Assumption 4. The right plot shows estimates of $\Psi(r,x)$ across simulated data sets when there are violations of Assumption 4, with an unmeasured confounder affecting both race and whether police use force against someone. Individual grey lines correspond to estimates from a single data set, while the blue line is the average across 100 repetitions of the same simulation.}
\end{figure} 

\section{Bounds for race and place estimand}

Here we show how one obtains bounds for $\Psi(r, x)$, the race and place estimand under a monotonicity assumption. Using the same derivations in Appendix A and assuming Assumptions 1 and 4, we can write
\begin{align*}
    E[Y(1) \mid R=r, X=x] &= E[Y(1) \mid R=r, X=x, M(1) = 1] \ P(M(1) = 1 \mid R=r, X=x) \\
    & \ + E[Y(1) \mid R=r, X=x, M(1) = 0] \ P(M(1) = 0 \mid R=r, X=x) \\
    &= E[Y(1,1) \mid R=r, X=x, M(1) = 1] \ P(M(1) = 1 \mid R=r, X=x) \\
    & \ + E[Y(1,0) \mid R=r, X=x, M(1) = 0] \ P(M(1) = 0 \mid R=r, X=x) \\
    &= E[Y(1,1) \mid R=r, X=x, M(1) = 1] \ P(M(1) = 1 \mid R=r, X=x) \\
    &= E[Y \mid R=r, X=x, M = 1, D=1] \ P(M(1) = 1 \mid R=r, X=x) c_1(x)
\end{align*}
This only relied on Assumption 1 and 4 holding, and did not rely on Assumption 5. Now suppose one is willing to assume monotonicity in the sense that
$$P(M(1) = 1 \mid R=r, X=x) \leq P(M(1) = 1 \mid R=r', X=x),$$
for any $r \geq r'$. Then we can see that for any value of $r \geq r_0$, we have that
\begin{align*}
    \Psi(r,x) &= \frac{E[Y(1) \mid R=r, X=x]}{E[Y(1) \mid R=r_0, X=x]} \\
    & = \frac{E[Y \mid R=r, X=x, M = 1, D=1] \ P(M(1) = 1 \mid R=r, X=x)c_1(x)}{E[Y \mid R=r_0, X=x, M = 1, D=1] \ P(M(1) = 1 \mid R=r_0, X=x)c_1(x)} \\
    & \leq \frac{E[Y \mid R=r, X=x, M = 1, D=1]}{E[Y \mid R=r_0, X=x, M = 1, D=1]} \\
    &=\Psi_{\text{na\"ive}}(r, x)
\end{align*}
Similar results hold for $r \leq r_0$, except the direction of the inequality will be flipped because the ratio 
$$\frac{P(M(1) = 1 \mid R=r, X=x)}{P(M(1) = 1 \mid R=r_0, X=x)}$$
is greater than or equal to 1, when $r \leq r_0$.

\section{Review of existing estimands and assumptions}

In this section, we briefly review assumptions that were covered in \cite{zhao2022note} in relation to estimating the causal risk ratio, along with a discussion of simple estimands that have been used previously in the literature targeting different quantities. A key assumption that effectively all analyses of this data require is what is referred to as a mandatory reporting assumption, which states
    \begin{align*}
    &(i) \text{ The administrative data includes all stops of civilians by the police.}\\
    &(ii) \ Y_{i}(d,0) = 0 \text{ for all } i \text{ and for } d \in \{0, 1\}
    \end{align*}
The validity of the first part of this assumption lies in the fact that reporting of all stops is mandated for all NYPD police officers. Additionally, the widespread presence of cameras and media/public interest in police brutality makes unrecorded uses of force increasingly unlikely. The second component of this assumption is very mild as it is reasonable to assume that police do not employ force against an individual unless they first detain that individual. 

A natural estimand targeting racial discrimination would be the average treatment effect (ATE), denoted here by $ E[Y(1)-Y(0)]$, which targets the overall extent to which Black individuals face a higher risk of police force compared to white individuals. This global causal effect is inherently difficult to estimate without strong assumptions \citep{knox2020administrative}, because we only observe data for which $M = 1$. Therefore, in studies with similar data sources \citep{gaebler2022causal}, focus has shifted to a local form of the ATE that focuses on the population with $M_i = 1$ given by $ATE_{M=1}$:
$$ \text{ATE}_{M=1}=E[Y(1)-Y(0)\mid M=1].$$
While local estimands such as these can be useful in many settings, in other situations global estimands are generally preferred \citep{swanson2014think}. This is especially true in our setting where discrimination in deciding to stop an individual may represent the largest component of discrimination, and ignoring it could lead to substantively different conclusions. Additionally, as noted by \cite{zhao2022note}, the local version of the ATE may even have the opposite sign compared with the global ATE. For this reason, we focus on global estimands in this article. While these come with difficulties in identification, we target estimands that aim to minimize questionable assumptions and incorporate readily available secondary data sources.  

Now we briefly detail the key ignorability assumption that the causal risk ratio, and many other estimands, are reliant upon for identification. This assumption says that there are no unmeasured confounders of the relationship between race and the outcomes of interest. In our setting, this assumption can be formalized as
    \begin{align*}
    &(i) \ Y (d,m) \perp D \mid X,M(d) \\
    &(ii) \ M(d) \perp D \mid X.
    \end{align*}
The justification for this assumption is that administrative datasets have evolved to incorporate numerous encounter attributes that encompass many features that correlate with both race and the use of force. It is still a strong and unverifiable assumption, however. For instance, if one race commits crimes at higher rates than another and this isn't accounted for by observed covariates, then part (ii) of the assumption would not hold. One can then perform sensitivity analysis to this assumption, and our sensitivity analysis framework in the manuscript would apply in this situation as well.  

\section{Derivations and additional results for sensitivity analysis framework}

In this section we provide all necessary derivations of quantities and formulas that were shown in Section 6 of the manuscript on sensitivity analysis. We also focused on $\xi(r, x)$ in the manuscript in relation to violations of Assumption 2(i), though we show what this quantity looks like under violations of the remaining assumptions in this section. Before discussing specifics of $\xi(r,x)$ under the different violations of assumptions, we can first derive the bias of the causal effect under violations of any of the assumptions. This is given below:
\begin{align*}
     E(\Psi(r, x) - \Psi^*(r, x)) &= E(\Psi(r, x)(1 - \xi(r, x))) \\
    &= Cov(\Psi(r, x), (1 - \xi(r, x))) + E(\Psi(r, x)) E(1 - \xi(r, x)) \\
    &= -Corr(\Psi(r, x), \xi(r, x)) \sqrt{Var(\Psi(r, x)) Var(\xi(r, x))} + E(\Psi(r, x)) (1 - E(\xi(r, x))).
\end{align*}
The form of $\xi(r, x)$ depends on which assumption is being violated. First, we can derive the form of this quantity under violations of Assumption 2(i) as was done in the manuscript. In Appendix A, we can see that our identification formula requires us to estimate
$$P[Y(1,1) = 1 \mid R=r, X=x, M(1) = 1]$$
If there are no confounders $U$ and we only need to adjust for $X$, then we can use that
$$P[Y(1,1) = 1 \mid R=r, X=x, M(1) = 1] = P(Y=1 \mid R=r, X=x, D=1, M=1),$$
which is what we have in the manuscript. If instead, we have confounding by $U$, then this quantity can be written as
\begin{align*}
    P[Y(1,1) = 1 \mid R=r, X=x, M(1) = 1] &= E_{U \mid R, X, M(1)} \bigg[ P[Y(1,1) = 1 \mid R=r, X=x, U = U, M(1) = 1] \bigg] \\
    &= E_{U \mid R, X, M(1)} \bigg[ P[Y = 1 \mid R=r, X=x, U = U, D=1, M = 1] \bigg]
\end{align*}
Now we can take the ratio of this quantity at $r$ and at $r_0$ as our estimand requires, which gives us
\begin{align*}
    &\frac{E_{U \mid R=r, X, M(1)} \bigg[ P[Y = 1 \mid R=r, X=x, U = U, D=1, M = 1] \bigg]}{E_{U \mid R=r_0, X, M(1)} \bigg[ P[Y = 1 \mid R=r_0, X=x, U = U, D=1, M = 1] \bigg]} \\
    =& \frac{E_{U \mid R=r, X, M(1)} \bigg[ \frac{f(U \mid R=r, M=1, D=1, X=x, Y=1) P(Y=1 \mid R=r, M=1, D=1, X=x)}{f(U \mid R=r, M=1, D=1, X=x)} \bigg]}{E_{U \mid R=r_0, X, M(1)} \bigg[ \frac{f(U \mid R=r_0, M=1, D=1, X=x, Y=1) P(Y=1 \mid R=r_0, M=1, D=1, X=x)}{f(U \mid R=r_0, M=1, D=1, X=x)} \bigg]} \\
    =& \frac{P(Y=1 \mid R=r, M=1, D=1, X=x)E_{U \mid R=r, X, M(1)} \bigg[ \frac{f(U \mid R=r, M=1, D=1, X=x, Y=1)}{f(U \mid R=r, M=1, D=1, X=x)} \bigg]}{P(Y=1 \mid R=r_0, M=1, D=1, X=x) E_{U \mid R=r_0, X, M(1)} \bigg[ \frac{f(U \mid R=r_0, M=1, D=1, X=x, Y=1)}{f(U \mid R=r_0, M=1, D=1, X=x)} \bigg]} \\
    =& W \left\{ \frac{E_{U \mid R=r, X, M(1)} \bigg[ \frac{f(U \mid R=r, M=1, D=1, X=x, Y=1)}{f(U \mid R=r, M=1, D=1, X=x)} \bigg]}{E_{U \mid R=r_0, X, M(1)} \bigg[ \frac{f(U \mid R=r_0, M=1, D=1, X=x, Y=1)}{f(U \mid R=r_0, M=1, D=1, X=x)} \bigg]} \right\},
\end{align*}
where $W$ is what we use if we only adjust for $X$ and ignore $U$. This shows the form of $\xi(r, x)$ as described in the manuscript. We discuss in the manuscript how three conditional dependencies have to hold in order to obtain bias and for this term to be $1$. As a reminder, these are:
\begin{enumerate}
    \item $U \indep  Y \mid X, R, M=1, D=1$
    \item (a) $U \indep  R \mid X, M=1, D=1, Y=1$, (b) $U \indep  R \mid X, M=1, D=1$, and (c) $U \indep  R \mid X, M(1) = 1$
    \item $f(U \mid R, X, M(1) = 1) = f(U \mid R, X, M=1, D=1)$
\end{enumerate}
Here we show that the third condition, which effectively shows that $U$ must also be conditionally associated with $D$, also leads to $\xi(r, x) = 1$. If $U$ is conditionally independent of $D$, then $$f(U \mid R=r, X=x, M(1) = 1) = f(U \mid R=r, X=x, M(1) = 1, D=1) = f(U \mid R=r, X=x, M = 1, D=1)$$
and therefore the expectation in the numerator (and similarly for the denominator) becomes
\begin{align*}
    & E_{U \mid R=r, X=x, M(1)=1} \bigg[ \frac{f(U \mid
     R=r, M=1, D=1, X=x, Y=1)}{f(U \mid R=r, M=1, D=1, X=x)} \bigg] \\ 
    = &E_{U \mid R=r, X=x, M=1, D=1} \bigg[ \frac{f(U \mid R=r, M=1, D=1, X=x, Y=1)}{f(U \mid R=r, M=1, D=1, X=x)} \bigg] \\
    =& \int_u \bigg[ \frac{f(U \mid R=r, M=1, D=1, X=x, Y=1)}{f(U \mid R=r, M=1, D=1, X=x)} \bigg] f(U \mid R=r, M=1, D=1, X=x) du \\
    =& \int_u f(U \mid R=r, M=1, D=1, X=x, Y=1) du \\
    =& 1
\end{align*}

If instead, we wanted to assess violations of Assumption 2(ii), then we can apply similar logic. Specifically, our identification formula requires us to estimate
$$P(M(1) = 1 \mid R=r, X=x).$$
If there are no unmeasured confounders $U$, and we only need to adjust for $X$, then we have that
$$P(M(1) = 1 \mid R=r, X=x) = P(M = 1 \mid R=r, X=x, D=1).$$
However, in the presence of unmeasured confounders, we have that
\begin{align*}
    P(M(1) = 1 \mid R=r, X=x) &= E_{U \mid R, X} \bigg[P(M(1) = 1 \mid R=r, X=x, U=U) \bigg] \\
    &=  E_{U \mid R, X} \bigg[P(M = 1 \mid R=r, X=x, U=U, D=1) \bigg]. \\
\end{align*}
Then, taking the ratio of this quantity at $r$ and $r_0$ and applying Bayes rule in the same manner as with Assumption 2 (i), we can see that
\begin{align*}
    \frac{ P(M(1) = 1 \mid R=r, X=x)}{ P(M(1) = 1 \mid R=r_0, X=x)} = \frac{P(M = 1 \mid R=r, X=x, D=1) E_{U \mid R=r, X} \bigg[ \frac{f(U \mid R=r, D=1, X=x, M=1)}{f(U \mid R=r, D=1, X=x)} \bigg]}{P(M = 1 \mid R=r_0, X=x, D=1) E_{U \mid R=r_0, X} \bigg[ \frac{f(U \mid R=r_0, D=1, X=x, M=1)}{f(U \mid R=r_0, D=1, X=x)} \bigg]},
\end{align*}
which shows that under violations of Assumption 2(ii), we have that
$$\xi(r, x) = \frac{E_{U \mid R=r, X} \bigg[ \frac{f(U \mid R=r, D=1, X=x, M=1)}{f(U \mid R=r, D=1, X=x)} \bigg]}{E_{U \mid R=r_0, X} \bigg[ \frac{f(U \mid R=r_0, D=1, X=x, M=1)}{f(U \mid R=r_0, D=1, X=x)} \bigg]}$$
Lastly, we need to derive the expression for $\widetilde{\xi}(r,z)$. As a reminder, when benchmarking we are using the following quantity as the true value of the estimand:
\begin{align*}
        \widetilde{\Psi}(r, z) &= \frac{E(Y(1) \mid R=r, X=x, U=u)}{E(Y(1) \mid R=r_0, X=x, U=u)}
\end{align*}
We can apply the identification formula from Appendix A to see that this is equivalent to
\begin{align*}
&\frac{E(Y \mid R=r, M=1, D=1, X=x, U=u)}{E(Y \mid R=r_0, M=1, D=1, X=x, U=u)} \frac{P(D=1 \mid R=r, M=1, X=x, U=u)}{P(D=1 \mid R=r_0, M=1, X=x, U=u)} \\
    &\times \frac{f(r \mid M=1, X=x, U=u)}{f(r_0 \mid M=1, X=x, U=u)} \frac{f(r_0 \mid D=1, X=x, U=u)}{f(r \mid D=1, X=x, U=u)}
\end{align*}
We again focus on violations of Assumption 2(i) for simplicity, but similar ideas hold for other assumptions. Focusing just on the outcome model portion of this identification formula, we can see that
\begin{align*}
& \frac{P(Y=1 \mid R=r, M=1, D=1, X=x, U = u)}{P(Y=1 \mid R=r_0, M=1, D=1, X=x, U = u)} \\
    &= \frac{f(u \mid R=r, M=1, D=1, X=x, Y=1) P(Y=1 \mid R=r, M=1, D=1, X=x)}{f(u \mid R=r, M=1, D=1, X=x)} \\
&\times \frac{f(u \mid R=r_0, M=1, D=1, X=x)}{f(u \mid R=r_0, M=1, D=1, X=x, Y=1) P(Y=1 \mid R=r_0, M=1, D=1, X=x)} \\
&= \frac{P(Y=1 \mid R=r, M=1, D=1, X=x)}{P(Y=1 \mid R=r_0, M=1, D=1, X=x)} \\
\times & \left( \frac{f(u \mid R=r, M=1, D=1, X=x, Y=1) f(u \mid R=r_0, M=1, D=1, X=x)}{f(u \mid R=r_0, M=1, D=1, X=x, Y=1) f(u \mid R=r, M=1, D=1, X=x)}  \right)
\end{align*}
and therefore, we obtain the form of $\widetilde{\xi}(r, z)$ seen in the manuscript. 

\section{Sensitivity analysis to marginalization weights} 

In Section 5 of the manuscript we discussed difficulties around marginalization, which centered around the fact that we could not marginalize over the covariate distribution of interest and are only able to marginalize over the covariate distribution in the $M=1$ population. Here we examine whether the race and place estimand marginalized over the conditional distribution of $X \mid M=1$ generalizes to the estimand marginalized over unconditional distribution $X$. To do so, we introduce a sensitivity analysis under a bounded density ratio assumption. Specifically, we assume the ratio of the unconditional probability function $f(x)$ to the conditional probability function $f(x\mid M=1)$ is bounded by a factor $\gamma>1$. This can be expressed as:
$$
\frac{1}{\gamma} \le w(x) \le \gamma \quad \text{for all } x.
$$
where we let $w(x) = f(x) / f(x \mid M=1)$. Here, $f(\cdot)$ denotes either the probability density function (PDF) for continuous covariates or the probability mass function (PMF) for discrete ones. This assumption limits the degree of covariate shift between the $M=1$ and overall population. To describe how to find the upper and lower bounds of the estimand marginalized over the full covariate distribution, we first consider the case in which $x$ follows a continuous distribution. For fixed $r$, the worst-case (lower) bound of race and place estimand solves the following linear program:
$$
\begin{aligned}
&\min_{w(\cdot)} &&\int_{\mathcal X}  \Psi(r,x)\,dF_{x} = \int_{\mathcal X}  w(x)\Psi(r,x)\,dF_{x\mid M=1}\\
&\mathrm{s.t.} && \frac{1}{\gamma} \le w(x) \le \gamma \quad \text{for almost every } x,\\
&&& \int_{\mathcal X} w(x)\,dF_{x\mid M=1} = 1
\end{aligned}
$$

Observe that a strictly feasible point $w\equiv 1$ satisfies Slater’s condition, so strong duality holds and the KKT conditions are necessary and sufficient. Let $\lambda\in\mathbb{R}$ be the multiplier for the normalization constraint, and $u(x)\ge0$, $v(x)\ge0$ be the multipliers for the lower and upper constraints. The Lagrangian function is:

$$
\begin{aligned}
\mathcal{L}(w, \lambda, u, v)
&= \int_{\mathcal{X}} \Psi(r,x) w(x) \, dF_{x\mid M=1}
+ \lambda \left( \int_{\mathcal{X}} w(x)\,dF_{x\mid M=1} - 1 \right)\\
&\quad + \int_{\mathcal{X}} u(x) \left( \frac{1}{\gamma} - w(x) \right) dF_{x\mid M=1}
+ \int_{\mathcal{X}} v(x) \left( w(x) - \gamma \right) dF_{x\mid M=1}.
\end{aligned}
$$

The first-order (stationarity) condition with respect to $w(x)$ is
$$
\Psi(r,x) + \lambda - u(x) + v(x) = 0,
$$
and the complementary slackness conditions are
$$
u(x)\big(w(x) - \tfrac{1}{\gamma}\big) = 0, \qquad v(x)\big(\gamma - w(x)\big) = 0.
$$

It follows that for almost every $x$, only one of the following can hold:

\begin{itemize}
    \item If $w(x) \in (\tfrac{1}{\gamma}, \gamma)$, then $u(x) = v(x) = 0$, which forces $\Psi(r,x)+ \lambda = 0$. 
    \item If $w(x) = \tfrac{1}{\gamma}$, then $u(x) \ge 0$, $v(x) = 0$, and so $u(x) = \Psi(r,x) + \lambda \ge 0$.
    \item If $w(x) = \gamma$, then $v(x) \ge 0$, $u(x) = 0$, and so $v(x) = -(\Psi(r,x) + \lambda) \ge 0$, i.e., $\Psi(r,x) + \lambda \le 0$.
\end{itemize}

Therefore, the optimal solution $w^*(x)$ must take the form

$$
w^*(x) =
\begin{cases}
\gamma, & \Psi(r,x) + \lambda < 0, \\
\tfrac{1}{\gamma}, & \Psi(r,x) + \lambda > 0, \\
\text{arbitrary in } [\tfrac{1}{\gamma}, \gamma], & \Psi(r,x) + \lambda = 0,
\end{cases}
$$

If $P_{X\mid M=1}\{\Psi(r,X)+\lambda=0\}=0$ (e.g.\ $\Psi(r,X)$ has a continuous distribution), then optimal weights $w^*$ almost surely takes only the two extremes of the box constraint and the value of $\lambda$ is uniquely determined by the normalization constraint $\int w^*(x)\,dF_{x\mid M=1} = 1$.

This bounding approach allows us to transparently quantify how sensitive our results are to differences between the observed covariate distribution (conditional on stops) and the unobserved target population covariate distribution. In our application, the function $\Psi(r,x)$ is discrete with support only on a finite region $\mathcal{B}\subset\mathbb{R}$, and therefore the normalization constraint becomes
$$
\sum_{j=1}^J \pi_j\, w_{x_j} \;=\; 1,\qquad 
\pi_j := P_{X\mid M=1}(X=x_j).
$$ 
In particular, outside the finite set of $\mathcal{B}$, the weights are forced exactly at the two box‐constraint
extremes. Even on $\mathcal{B}$, one may assign extreme values to as many points in $\mathcal{X}_0$ as possible, leaving just a single point to be adjusted in order to satisfy the normalization constraint, without altering the objective value. We work with the empirical distribution of the observed covariates $\{x_i\}_{i=1}^n$ from the sample where $M_i=1$. Let $\Psi_i=\Psi(r,x_i)$. The optimization problem becomes finding weights $\{w_i\}_{i=1}^n$ to minimize $\frac{1}{n}\sum_{i=1}^n w_i\Psi_i$ subject to $1/\gamma\le w_i\le\gamma$ and $\frac{1}{n}\sum_{i=1}^n w_i=1$.
To construct the optimal solution, we first sort the observations such that $\Psi_{(1)}\le\Psi_{(2)}\le\cdots\le\Psi_{(n)}$. The optimal weights $w_{(i)}^*$ are as follows: 

$$
w_{(i)}^* =
\begin{cases}
\gamma, & i \le k, \\
\omega^\star, & i = k+1, \\
1/\gamma, & i \ge k+2,
\end{cases}
$$
where the integer $k \in \{0, 1, \dots, n-1\}$ is chosen such that the pivotal weight $\omega^\star$ falls within its prescribed bounds. The value of $\omega^\star$ is determined by the normalization constraint $\sum w_{(i)}^* = n$:
$$
k\gamma + \omega^\star + (n-k-1)\frac{1}{\gamma} = n \implies \omega^\star = n - k\gamma - (n-k-1)\frac{1}{\gamma}.
$$
The correct value of $k$ is the one that ensures $1/\gamma\le\omega^\star\le\gamma$. Such a $k$ is guaranteed to exist and is unique. This structure provides a simple, computationally efficient method for finding the worst-case bounds. 

\begin{figure}[h]
 \centering
    \includegraphics[width=0.85\linewidth]{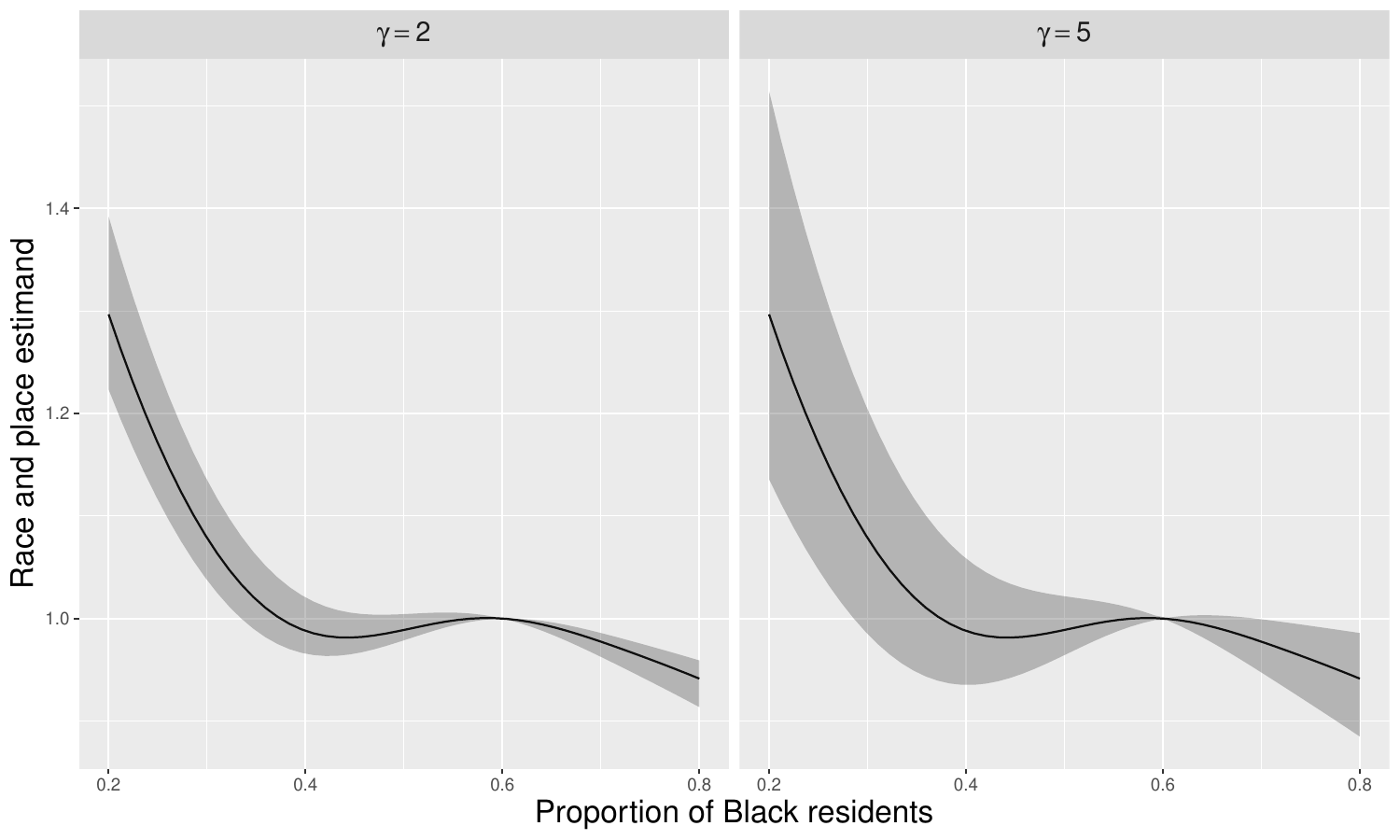} 
    \caption{Upper and lower bounds under the proposed sensitivity analysis framework under varying levels of $\gamma$.}
    \label{fig:AnalysisMarginal}
\end{figure}

We then apply these bounds to our NYC policing application, and the results can be found in Figure \ref{fig:AnalysisMarginal}. We explore values of $\gamma \in \{2, 5\}$, which both correspond to rather large discrepancies (particularly $\gamma = 5$) between the $M=1$ and overall population's covariate distributions.  We find that the overall conclusions remain the same showing that our conclusions about race and place policing are robust to the weights used in marginalization. When $\gamma = 2$, the upper and lower bounds are quite close to the estimated effects from the $M=1$ population showing a very similar result. The bounds are wider in the extreme case where $\gamma = 5$, but the overall trend is still negative, showing the existence of race and place policing. 

\section{Details of formal benchmarking procedure}

Here we describe a two step procedure to doing sensitivity analysis. The first stage is relatively straightforward and involves benchmarking partial $R^2$ values using the methodology developed in \cite{cinelli2020making}. The second stage examines how large our sensitivity parameters can be under the proposed partial $R^2$ values obtained from formal benchmarking. In the end, our goal is to construct a synthetic unobserved confounder, denoted $\tilde U$, that has the desired partial $R^2$ values with the observed variables.


Specifically, we construct a synthetic confounder, $\tilde U$, under the constraint that it is orthogonal to the observed covariates $X$, i.e., $\tilde U \perp \mathrm{span}\{1,X\}$. Here, we use the notation $R_{\mid X}$ to denote the residual from the OLS regression of a variable $R$ on a set of variables $X$. We aim to have $\tilde U$ match two key partial $R^2$ targets that affect our sensitivity parameters:

\begin{align*}
&\text{(1)}\quad
R^2_{YU\mid RX}
:= \operatorname{Corr}^2(Y_{\mid (R,X)},\, U_{\mid (R,X)}) \\[4pt]
&\text{(2)}\quad
R^2_{RU\mid X}
:= \operatorname{Corr}^2(R_{\mid X},\,  U_{\mid X})
\end{align*}

We also explore all possible signs leading to the same partial $R^2$: $\operatorname{sign}\!\big(\operatorname{Corr}(Y,U \mid R,X)\big),\operatorname{sign}\!\big(\operatorname{Corr}(R,U \mid X)\big)\in\{-1,+1\}$, which encode the direction of the pairwise associations. Let
$$
c   = \operatorname{sign}\!\big(\operatorname{Corr}(Y,U \mid R,X)\big)\sqrt{R^2_{YU\mid RX}}=\operatorname{Corr}(Y,U \mid R,X),\quad
a   = \operatorname{sign}\!\big(\operatorname{Corr}(R,U \mid X)\big)\sqrt{R^2_{RU\mid X}} =\operatorname{Corr}(R,U \mid X),\quad
$$
The construction of $\tilde U$ is based on a linear combination of components that are, by design, orthogonal to the column space of $X$. The final variable takes the form:
$$
\tilde U = u_\perp + \alpha_R R_{\mid X}
$$
where each component is chosen to meet a specific partial $R^2$ target without disturbing the other. First, we have that $u_{\perp}\in\mathrm{span}\{1,X,R\}^\perp$ is correlated with the residual for $Y$, i.e. $Y|_{R,X}$ so that $R^2_{YU\mid RX}$ equals its target value. Because $u_{\perp}$ is orthogonal to $(X,R)$, adding $u_{\perp}$ affects $R^2_{RU\mid X}$ only by affecting the variance of $\tilde U$ and it is orthogonal to $X$.
Second, the component $\alpha_R R_{\mid X}$ is tuned to meet the $R^2_{RU\mid X}$ target. Since $R_{\mid X}$ is a linear combination of $R$ and $X$, this component vanishes when computing the residual $\tilde U_{\mid (R,X)}$, thereby leaving the $R^2_{YU\mid RX}$ target unaffected.
Since both $u_\perp$ and $R_{\mid X}$ are orthogonal to $X$, their sum $\tilde U$ also satisfies the orthogonality constraint $\tilde U \perp X$.

\paragraph{Algorithm.}
The algorithm begins by computing two key OLS residuals that form the basis of our construction:
$$
e_Y := Y_{\mid(R,X)} \quad \text{and} \quad R_{\mid X}.
$$
The residual for $Y$ is then standardized for use in the first step: $\tilde e_Y = e_Y/\sigma(e_Y)$.

\paragraph{Step 1 (Targeting $R^2_{YU\mid RX}$.)}
$$
u_\perp \;=\; c\,\tilde e_Y \;+\; \sqrt{1-c^2}\,z.
$$
Here let $z$ be any unit vector orthogonal to $\mathrm{span}\{1,X,R,e_Y\}$. Then $u_\perp\in\mathrm{span}\{1,X,R\}^\perp$ and 
$\mathrm{Corr}\!\big(Y_{\mid(R,X)},u_\perp\big)=c$, hence 
$\mathrm{Corr}^2\!\big(Y_{\mid(R,X)},u_\perp\big)=R^2_{YU\mid RX}$.  

\paragraph{Step 2 (Targeting $R^2_{RU\mid X}$).}
$$
\alpha_R \;=\; \frac{a}{\sqrt{1-a^2}}\cdot\frac{\sigma(u_\perp)}{\sigma(R_{\mid X})}.
$$
Since $X^\top \beta \in \operatorname{col}(X)$, we have $\tilde U_{\mid X}=u_\perp+\alpha_R R_{\mid X}$. By construction $u_\perp\perp R_{\mid X}$.  
Write $s_u=\sigma(u_\perp)$ and $s_r=\sigma(R_{\mid X})$.  
Then
$$
\mathrm{Cov}(\tilde U_{\mid X},R_{\mid X})
=\mathrm{Cov}(u_\perp+\alpha_R R_{\mid X},R_{\mid X})
=\alpha_R s_r^2,
$$
and
$$
\sigma(\tilde U_{\mid X})=\sqrt{\sigma(u_\perp)^2+\alpha_R^2\,\sigma(R_{\mid X})^2}
=\sqrt{s_u^2+\alpha_R^2 s_r^2}.
$$
Hence,
$$
\mathrm{Corr}(\tilde U_{\mid X},R_{\mid X})
=\frac{\alpha_R s_r^2}{\sigma(\tilde U_{\mid X})\,s_r}
=\frac{\alpha_R s_r}{\sqrt{s_u^2+\alpha_R^2 s_r^2}}.
$$
With
$$
\alpha_R=\frac{a}{\sqrt{1-a^2}}\cdot\frac{s_u}{s_r},
$$
we get
$$
\mathrm{Corr}(\tilde U_{\mid X},R_{\mid X})
=\frac{\tfrac{a}{\sqrt{1-a^2}}\,s_u}{\sqrt{s_u^2+\tfrac{a^2}{1-a^2}\,s_u^2}}
=\frac{\tfrac{a}{\sqrt{1-a^2}}\,s_u}{\tfrac{s_u}{\sqrt{1-a^2}}}
=a,
$$
and therefore
$$
\mathrm{Corr}^2(\tilde U_{\mid X},R_{\mid X})=R^2_{RU\mid X}.
$$

\paragraph{Signs and subsequent analysis}
The directions $c$ and $a$ depend on the signs of the target partial correlations and can affect the resulting sensitivity parameters and bias. To address this, our procedure is repeated for all possible sign combinations, which allows us to explore the full range of potential impacts the synthetic confounder could have, given its specified strength. Once a synthetic confounder $\tilde U$ has been constructed for a given set of targets and a specific sign combination, the subsequent step is to quantify its impact on the key parameters in our sensitivity framework. This is achieved by incorporating $\tilde U$ as an additional covariate into the models that describe the full data generating process. Using the outcome model as an example, we compare the ratio estimated from the reduced model with that from the confounder-augmented full model. The reduced model is:
\begin{equation} 
\text{Outcome}_{\text{reduced}}(r, r_0, x) = \frac{E[Y \mid R=r, M=1, D=1, X=x]}{E[Y \mid R=r_0, M=1, D=1, X=x]}.
\end{equation}
The full model is conditioned on $\tilde U$:
\begin{equation} 
\text{Outcome}_{\text{full}}(r, r_0, x, u) = \frac{E[Y \mid R=r, M=1, D=1, X=x, U=\tilde{U}]}{E[Y \mid R=r_0, M=1, D=1, X=x, U=\tilde{U}]}.
\end{equation}
The ratio of these two quantities forms a component of our sensitivity metric, $\xi(r,x) = \text{RR}_{\text{full}} / \text{RR}_{\text{reduced}}$, which quantifies the bias amplification due to $\tilde U$. A similar procedure can be done simultaneousely for other models used in our identification formula such as the model for $D$ and the model for $R$. We can then examine the mean and variance of the resulting $\xi(r,x)$ and use these as our benchmark values in a sensitivity analysis.

\section{Simulation highlighting pitfalls of informal benchmarking}

In this section we present a simple simulation study that highlights how informal benchmarking can lead to misleading findings within a sensitivity analysis even if the researcher has correct prior knowledge about the relative strength of association for an unmeasured variable compared with an observed one used in benchmarking. Using a similar idea as in \cite{cinelli2020making}, we simulate an observed covariate and an unobserved covariate to be the exact same in every way, and show that informal benchmarking underestimates the magnitude of sensitivity parameters, but our formal benchmarking procedure described in the previous section addresses this concern. Specifically, we simulate $n=10^5$ observations from the following process:
$$
\begin{aligned}
&X\sim\mathcal N(0,1),\quad U\sim\mathcal N(0,1),\quad
R=X+U+\varepsilon_R,\;\varepsilon_R\sim\mathcal N(0,1),\\
&D\sim\mathrm{Bernoulli}\!\big(\mathrm{expit}(X+ U+0.5\,R)\big),\\
&M\sim\mathrm{Bernoulli}\!\big(\mathrm{expit}(X+ U+0.5\,R+0.5\,D)\big),\\
&Y\sim\mathrm{Bernoulli}\!\big(\mathrm{expit}(X-U+0.5\,R+0.5\,D+0.5\,M)\big).
\end{aligned}
$$
Informal benchmarking is performed in this case by taking the ratio of predictions from models that 1) do not include either $X$ or $U$ and 2) models that include $X$ but not $U$. We can refer to the bias multiplier values $\xi(r,x)$ obtained from this process as $\xi_X$ since they are obtained by including and excluding $X$ from the models, as is standard in benchmarking. Similarly, we can take the ratio of predictions from models that 1) do not include $U$ and 2) models that include both $X$ and $U$. We can refer to these values by $\xi_U$, and these are the values of interest to us. If the researcher correctly assumes that the impact of $X$ on $(D, R, Y)$ is the same in magnitude as the impact of $U$ on $(D, R, Y)$, then informal benchmarking would amount to assuming $E(\xi_X) = E(\xi_U)$ and $sd(\xi_X) = sd(\xi_U)$. Unfortunately, however, we obtain the following result in simulation:
$$
E[\xi_U] =  1.071,\;\; \mathrm{sd}(\xi_U)= 0.063,
\qquad
\mathbb E[\xi_X]= 0.951,\;\; \mathrm{sd}(\xi_X)= 0.012.
$$
We see that the informal benchmarking procedure drastically underestimates the variability in $\xi_U$, and somewhat underestimates how far $E(\xi_U)$ is from 1. We have seen empirically in other simulated examples as well that such informal benchmarking can drastically underestimate both sensitivity parameters. We can apply our formal benchmarking procedure described in the previous section, however, to see if it addresses the issues inherent to informal benchmarking. Doing so, we obtain the following estimates of the sensitivity parameters
$$
\widehat{E}[\xi_U] =  1.078,\;\; \widehat{\mathrm{sd}}(\xi_U)= 0.065.
$$
This shows that the formal benchmarking procedure was successful in capturing the correct sensitivity parameters in this case. For this reason, we proceed with using this formal benchmarking procedure in all benchmarking analyses done in the NYC policing analysis.

\section{Race and place results using more recent data only}

\begin{figure}[h]
 \centering
    \includegraphics[width=0.65\linewidth]{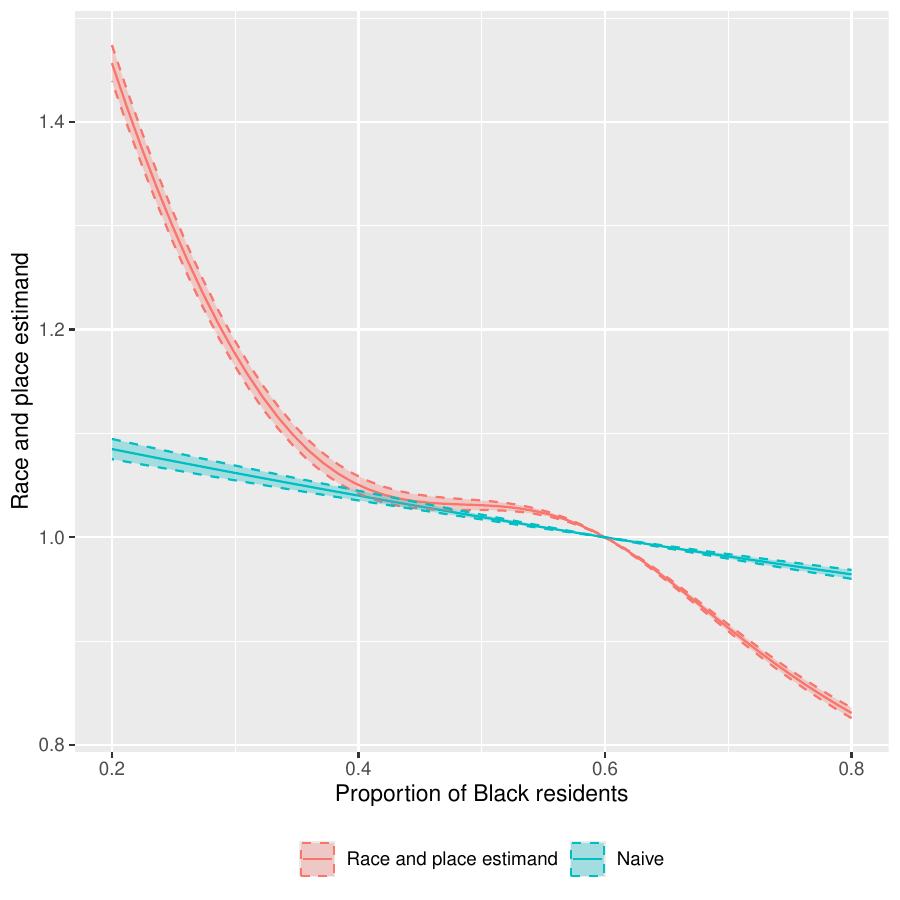} 
    \caption{Estimates and 95\% confidence intervals for $\overline{\Psi}(r)$ and the na\"ive estimator of this quantity obtained by averaging $\Psi_{\text{na\"ive}}(r, x)$ over the distribution of $X$ when only examining data from 2012 and 2013}
    \label{fig:Analysis2012results}
\end{figure}

In this section, we perform the same analysis as the main race and place analysis, which focuses on estimating $\overline{\Psi}(r)$, except we restrict to police stops that occurred in 2012 and 2013. The analysis of the main manuscript focused on all years of available stop data, which included 2003 to 2013. We focus on more recent years here as they more closely align with the timing of our cell phone mobility data, which was collected from the year 2019. If mobility patterns change drastically over short time periods, then this analysis should be more robust against bias that occurs due to incorrect estimates of mobility patterns across NYC.  We calculate both the na\"ive estimand, as well as the estimand of interest $\overline{\Psi}(r)$. The results for both, along with pointwise 95\% confidence intervals can be found in Figure \ref{fig:Analysis2012results}. We see that the results are fairly similar to those obtained in the manuscript. The na\"ive estimand is estimated to be relatively flat, although the slope is negative here instead of positive as in the manuscript. Note, however, that this estimand is completely unaffected by mobility data, and therefore any changes are simply due to random variation or differences of effects across time. The estimate for $\overline{\Psi}(r)$ is similar to the one found in the manuscript with a steep, negative slope indicating that Black people are more likely to have force used against them if they are in predominantly white neighborhoods. This effect is slightly more pronounced in this analysis with a slightly more negative slope, but overall the conclusions are very similar.  

\section{Analysis of mobility patterns over time with LODES data}

\begin{figure}[h]
 \centering
    \includegraphics[width=0.45\linewidth]{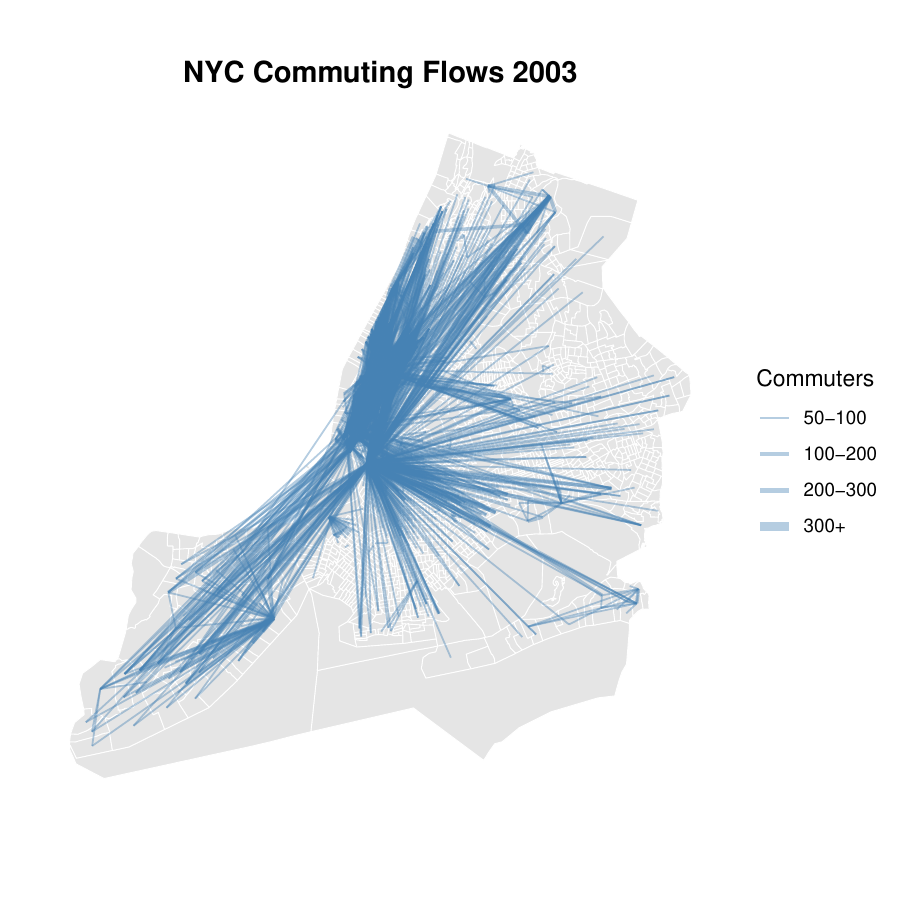} 
    \includegraphics[width=0.45\linewidth]{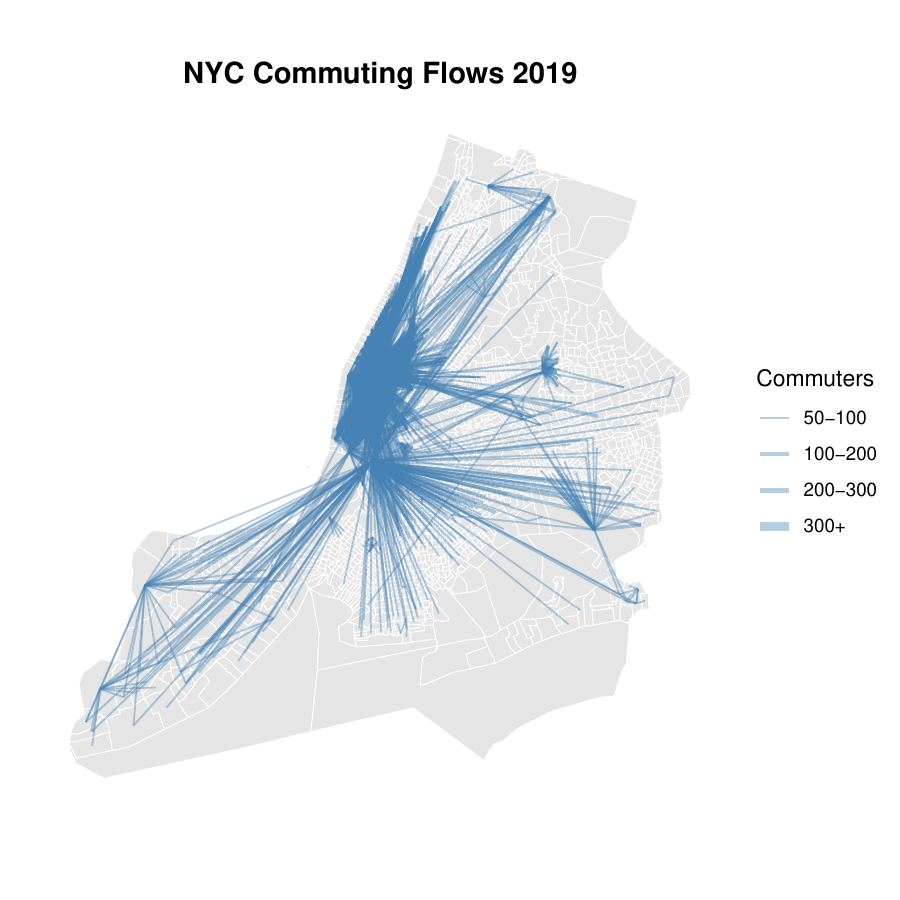} 
    \caption{Employment-based Commuting patterns in the years 2003 and 2019 taken from the U.S. Census LODES data}
    \label{fig:LodesResults}
\end{figure}

In this section, we analyze a secondary source of mobility data that is available on a yearly basis, which we use to study the degree to which mobility patterns change over time in NYC. The United States Census LODES data (Longitudinal Employer-Household Dynamics Origin-Destination Employment Statistics) provides information at the Census tract level about where individuals both live and work, which allows us to study commuting patterns for individuals who commute to work. This data source provides the number of individuals in any particular Census tract who work in any other Census tract. We do not use this data source as our main source of information for mobility as it only provides information on commuting for employment and does not provide information on the amount of time spent in each Census tract. Nonetheless, it is available on a yearly basis and therefore we can use it to assess the extent to which mobility patterns have changed over time. This is important to assess as our cell phone mobility data was collected in the year 2019, which differs from the years of our study (2003-2013), and it is important to understand how representative (or not) our mobility data is. To study this, Figure \ref{fig:LodesResults} shows mobility patterns from the LODES data in both the first year of our study (2003) and the year from which our mobility data comes from (2019). To avoid clutter in the figures, we restrict to all pairs of locations that have at least 50 commuters, and therefore we are focusing on the largest and most important commuting routes. We see that while there are some differences across these two years, general mobility patterns are largely the same within NYC over this time frame. Mobility has decreased slightly, at least for employment, from 2003 to 2019, and there are only mild changes in the overall patterns. Overall, this shows that our cell phone mobility data is likely a reasonable approximation to general mobility patterns over the course of our study. While it would be ideal to have data over the exact time frame of our study, we believe that using this source of mobility data is closer to the true mobility patterns of individuals in NYC than ignoring mobility altogether.  

\section{Details for interpretability simulations}

This section introduces the data generating processes for the two simulations in the main text that show how the sensitivity parameters vary as a function of specific regression parameters. These simulations are designed to illustrate the interpretation and robustness of our proposed estimands under different scenarios where our core assumptions are violated.

\paragraph{Data-generating processes for the confounding simulation}

The first simulation investigates a scenario with a binary unmeasured confounder $U$ that directly affects both the mediator $M$ and the outcome $Y$. The strength of the
confounding is controlled by a parameter $a \in [0,1]$. For each of the $n = 500,000$ units, the data are generated as follows:

\begin{align*}
U &\sim \mathrm{Bernoulli}(0.5) \\
X &\sim \mathcal{N}(0, 1) \\
\varepsilon_R &\sim \mathcal{N}(0, 1) \\
R' &= a_1 \cdot a \cdot U + X + \varepsilon_R \\
R &= 0.1 + \frac{R' - \min(R')}{1.2 \times (\max(R') - \min(R'))} \quad \\
D &\sim \mathrm{Bernoulli}\big(\mathrm{plogis}(a_2 \cdot a \cdot U + X)\big) \\
M &\sim \mathrm{Bernoulli}\big(\mathrm{plogis}(a_3 \cdot a \cdot U + X + 0.5D + R + R \cdot X)\big) \\
Y &\sim \mathrm{Bernoulli}\big(\mathrm{plogis}(-0.6 + a_4 \cdot a \cdot U + R + 0.5X + 0.5M + 0.5D + R \cdot X)\big)
\end{align*}

where $\mathrm{plogis}(v) = 1/(1+e^{-v})$ is the inverse logit function. We vary the parameters $a_1, a_2, a_3, a_4 \in \{-1, +1\}$ to simulate all possible directions of the unobserved confounding effects.

\paragraph{Data-generating processes for violations of Assumption 3}

This simulation highlights a scenario where violations of our assumptions arise due to differential impacts of $R$ and $X$ on both $M$ and $Y$, which leads to violations of Assumption 3. These effects are varied through interaction terms, whose strength is controlled by a parameter $a \in [0,0.5]$. For each of the $n = 500,000$ units, the data are generated as follows:
\begin{align*}
X &\sim \mathcal{N}(0, 1) \\
R &\sim \mathcal{N}(-0.5X, 1) \\
D &\sim \mathrm{Bernoulli}\big(\mathrm{plogis}(0.3X - 0.3R)\big) \\
M &\sim \mathrm{Bernoulli}\big(\mathrm{plogis}(0.3X - 0.3R + 0.3D + a_1 \cdot a \cdot D \cdot X + a_2 \cdot a \cdot R \cdot X)\big) \\
Y &\sim \mathrm{Bernoulli}\big(\mathrm{plogis}(-0.6 + 0.5R - 0.5X + 0.5M + 0.5D + a_3 \cdot a \cdot R \cdot X)\big)
\end{align*}
Again, we vary the parameters $a_1, a_2, a_3 \in \{-1, +1\}$ to simulate all possible directions of the unobserved confounding effects.

\end{document}